\documentclass[
superscriptaddress,
preprint,
 amsmath,amssymb,
 aps,
prb,
titlepage,eqsecnum]{revtex4-2}
\usepackage[mathscr]{euscript}
\usepackage{bm}
\usepackage{exscale}
\usepackage{textcomp}
\usepackage{comment}
\usepackage{slashed}
\usepackage{tabularx}
\usepackage{wrapfig}
\usepackage{makecell}
\usepackage{amsmath,amsfonts,amsthm,amsbsy,mathtools} 
\usepackage{graphicx} 
\usepackage{bbold} 
\usepackage{physics}
\usepackage{hyperref} 
\usepackage[dvipsnames]{xcolor} 
\usepackage{enumitem} 
\usepackage{pst-all}
\usepackage[caption=false,
    labelformat=simple,
    listofformat=subsimple]{subfig}
\usepackage{makecell}
\usepackage[normalem]{ulem} 

\definecolor{MathematicaPink}{RGB}{255, 128, 128}
\definecolor{MathematicaCyan}{RGB}{0, 179, 255}
\definecolor{MathematicaPurple}{RGB}{128, 0, 128}
\makeatletter
\renewcommand{\figurename}{Figure}
\renewcommand{\fnum@figure}{\textbf{\figurename~\thefigure}}

\renewcommand{\tablename}{Table}
\renewcommand{\fnum@table}{\textbf{\tablename~\thetable}}
\makeatother
\newcommand{\id}{\mathbb{1}}

\def\sgn{\text{sgn}}
\def\hc{\text{H.c.}}
\newcommand{\pbr}[1]{\left(#1\right)}
\newcommand{\sbr}[1]{\left[#1\right]}
\newcommand{\cbr}[1]{\left\{#1\right\}}

\def\Z{\mathbb{Z}}

\def\L{\mathcal{L}}
\def\G{\mathcal{G}}
\def\H{\mathcal{H}}
\def\T{\mathcal{T}}
\def\A{\mathcal{A}}
\def\B{\mathcal{B}}
\def\C{\mathcal{C}}
\def\F{\mathcal{F}}
\def\M{\mathcal{M}}
\def\D{\mathcal{D}}
\def\O{\mathcal{O}}
\def\P{\mathcal{P}}
\def\S{\mathcal{S}}
\def\U{\mathcal{U}}

\def\Zp{\mathcal{Z}} 

\def\da{\dagger}

\def\rd{{\rm d}}
\def\pa{\partial}

\def\ra{\rightarrow}

\def\sl{\slashed}
\def\jb{\overline{j}}

\def\al{\alpha}
\def\be{\beta}
\def\de{\delta}
\def\De{\Delta}
\def\ep{\epsilon}
\def\ga{\gamma}
\def\Ga{\Gamma}
\def\la{\lambda}
\def\La{\Lambda}
\def\ka{\kappa}

\def\pa{\partial}

\def\vep{\varepsilon}
\def\vp{\varphi}
\def\vt{\vartheta}
\def\si{\sigma}
\def\Si{\Sigma}
\def\th{\theta}
\def\Th{\Theta}

\newcommand{\JC}[1]{{\color{Green}[JC: #1]}}
\newcommand{\RS}[1]{{\color{red}[RS: #1]}}

\hypersetup{
    unicode=false,          
    pdftoolbar=true,        
    pdfmenubar=true,        
    pdffitwindow=true,     
    pdfstartview={FitH},    
    pdftitle={ZM-in-FQH-SC},    
    pdfauthor={},     
    pdfsubject={},   
    pdfcreator={},   
    pdfproducer={}, 
    pdfkeywords={zero mode} {non-Abelian anyons} {symmetry fractionalization} {symmetry defects}, 
    pdfnewwindow=true,      
    colorlinks=true,       
    linkcolor=MathematicaPurple, 
    citecolor=blue,        
    filecolor=MathematicaPink,      
    urlcolor=MathematicaCyan      
}

\begin{document}

\title{Zero Mode Operators and Charge-Conjugation Defects in Non-Abelian Fractional Quantum Hall-Superconductor Heterostructures}
\author{Junyi Cao}
\author{Eduardo Fradkin}
\affiliation{Department of Physics and The Anthony J. Leggett Institute for Condensed Matter Theory, University of Illinois Urbana-Champaign, 1110 West Green Street, Urbana, IL 61801, USA}
\begin{abstract}
    We study zero modes bound to charge-conjugation defects in fractional quantum Hall-superconductor (FQH-SC) heterostructures built from Abelian and non-Abelian FQH states. Although the associated defects are well described by $G$-crossed braided tensor categories, a local operator description of their zero modes is not generally known. We generalize the bulk-to-boundary construction of localized zero mode operators to FQH-SCs with non-Abelian parent topological orders, mapping bulk Wilson operators in the folded theory to localized zero mode operators in the one-dimensional FQH-SC theory. This construction distinguishes zero mode anyons from the local zero mode operators that act on the defect Hilbert space and yields a generalized zero mode algebra. It also determines the fusion-channel dependence of tunneling: tunneling of zero mode operators probes bulk monodromy, while direct zero mode anyon transfer probes $G$-crossed associativity data. Applying this framework, we recover the $\Z_{2m}$ parafermion zero mode structure of fermionic Laughlin FQH-SCs at filling $\nu=1/m$, obtain a $\Z_2^{(n)}\times\Z_{N}$ structure for generalized Moore–Read states at filling $\nu=1/m$, with $N=2m$ for fermionic states and $N=m$ for bosonic states, and find a $\text{Fib}\times\Z_N$ structure for the $k=3$ Read–Rezayi state at filling $\nu=3/(3M+2)$, with $N=2(3M+2)$ for fermionic states and $N=(3M+2)/2$ for bosonic states. In the Read–Rezayi case, zero mode operator tunneling and direct zero mode anyon transfer yield distinct universal amplitude ratios between the neutral fusion channels, providing a characteristic signature of the underlying Fibonacci defect structure.
\end{abstract}
\date{\today}
\maketitle

\tableofcontents
\section{Introduction}\label{sec:intro}
Topological quantum computation uses the degenerate states associated with several non-Abelian anyons to store quantum information nonlocally. When the anyons are spatially separated, local perturbations cannot distinguish these states, providing protection against decoherence. Braiding the anyons implements unitary transformations within this space of protected states and thereby provides a route toward fault-tolerant quantum computation \cite{freedman-2000,kitaev-2003,nayak-2008,simon-2023}. 

Non-Abelian fusion rules and braiding statistics can also arise from \emph{twist defects} in topological states of matter \cite{bombin-2010,barkeshli-2012,you-2012,you-2013,barkeshli-2013-genon,barkeshli-2013,barkeshli-2019}. Throughout this paper, we refer to twist defects simply as defects. Unlike anyons which are dynamical excitations, these defects are imposed externally, and thereby can be better controlled through the device geometry. These defects exhibit well-defined fusion relations, and their effective braiding operations can be implemented through topological charge measurements or tunable interactions \cite{bonderson-2008,bonderson-2009,bonderson-2013}. Another advantage of these twist defects is that they can support non-Abelian fusion spaces even with an Abelian parent topological order (TO).

Early proposals considered lattice dislocations that permute anyon types \cite{bombin-2010,barkeshli-2012,you-2012}. Related constructions include layer-exchange defects, known as genons, in multilayer topological states \cite{barkeshli-2012,barkeshli-2013-genon,barkeshli-2014}, and domain walls between differently gapped edges \cite{barkeshli-2013,clarke-2013,lindner-2012,cheng-2012,barkeshli-2013-2,cong-2017}. We focus on fractional quantum Hall–superconductor (FQH-SC) heterostructures, where the superconducting geometry fixes the defect positions and allows their couplings to be controlled. Josephson and quasiparticle tunneling measurements provide probes of the associated fusion channels \cite{clarke-2013,lindner-2012,cheng-2012,clarke-2014,cao-2026-l}. This setting also allows localized zero mode operators and their couplings to be constructed directly from the parent edge theory.

The defects carry non-Abelian fusion and braiding data, while their bound zero modes are described by localized operators acting on their fusion space, known as the \emph{defect Hilbert space}. Coupling zero modes at different defects can lift the degeneracy between fusion channels within the defect Hilbert space \cite{kitaev-2001}, providing a probe of the nonlocally encoded topological charge. Such couplings can be used for quantum-information processing and experimental characterization of topological phases. In the Abelian examples, these endpoint defects bind parafermion zero modes (PZMs) \cite{clarke-2013,lindner-2012,cheng-2012}. 

The parafermion operators appearing in these constructions originated from one-dimensional $\Z_N$ clock models, where they arise through a generalized Jordan–Wigner transformation \cite{fradkin-1980}. These models can have critical points described by $(1+1)$-dimensional $\Z_N$ parafermion conformal field theories (CFTs) \cite{zamolodchikov-1985}. The operator product expansions of the primary fields encode fusion rules analogous to those of bulk anyons. For $N=3$, the parafermion CFT includes a sector with Fibonacci fusion rules. The corresponding Fibonacci anyons \cite{read-1999} support universal topological quantum computation (TQC) \cite{freedman-2000}. 

The theory and physical realizations of defects and their bound zero modes have been developed through extensive work \cite{barkeshli-2012,fendley-2012,barkeshli-2013-genon,barkeshli-2013,barkeshli-2013-2,barkeshli-2014,clarke-2014,alicea-2016,cong-2017-2}. In particular, the fusion and braiding properties of these twist defects are captured by the algebraic theory of symmetry defects which extends that of anyons to include anyonic symmetries (see App. \ref{subapp:Gcross} for technical details). However, a corresponding local operator description has been established only for Abelian parent topological orders, where bulk Wilson operators---representing anyon world lines in the bulk---are related to localized parafermion operators at domain walls between gapped boundaries \cite{barkeshli-2013}. 
However, a general construction of localized zero mode operators for non-Abelian parent topological orders remains lacking. Without such a construction, the coupling between nearby defects---and hence the fusion-channel-dependent tunneling response---cannot be formulated directly in the one-dimensional FQH-SC theory.


In this work, we generalize the construction of localized zero mode operators to FQH-SCs with non-Abelian parent topological orders. Building on the established interpretation of an FQH-SC segment as a charge-conjugation defect line \cite{barkeshli-2013,khan-2017,barkeshli-2019}, we distinguish between the topological charges that can be absorbed or emitted by a defect without creating an excitation, which we call \emph{zero mode anyons}, and their local realization in the one-dimensional FQH-SC theory, which we call \emph{zero mode operators}. We map bulk Wilson operators to localized zero mode operators and express the zero mode operators in terms of primary fields of the edge theory corresponding to the parent state. This mapping translates the bulk Wilson line algebra into a generalized zero mode algebra acting on the defect fusion space. This distinction separates two physical processes: tunneling of the zero mode operators and direct transfer of a zero mode anyon between two defects. We determine how these processes split the energies of the defect fusion channels and contribute to the Josephson response.

\begin{table}
    \centering
    \begin{tabular}{||c|c|c||}
        \hline 
        Parent FQH state&\makecell{Zero mode\\structure} &Characteristic tunneling signature\\
        \hline
        $\nu=1/m$ fermionic Laughlin ($m$ odd)& $\Z_{2m}$& $4m\pi$-periodic Josephson effect\\
        $\nu=1/m$ bosonic Laughlin ($m$ even)& $\Z_{m/2}$&$m\pi$-periodic Josephson effect\\
        $\nu=2/5$ Jain/Halperin $(332)$& $\Z_{5}$&$10\pi$-periodic Josephson effect\\
        $\nu=2/3$ Jain/Halperin $(1,1,2)$&$\Z_{3}$&$6\pi$-periodic Josephson effect\\
        $\nu=3/5$ hole-conjugated Jain/Halperin & $\Z_{10}$&$20\pi$-periodic Josephson effect\\
        $\nu=1/2$ Halperin $(331)$&$\Z_{4}$&$8\pi$-periodic Josephson effect\\
        \hline
        $\nu=1/m$ fermionic Moore-Read ($m$ even)& $\Z_2^{(n)}\times\Z_{2m}$&equal neutral-channel magnitudes\\
        $\nu=1/m$ bosonic Moore-Read ($m$ odd)& $\Z_2^{(n)}\times\Z_{m}$&and a relative $\pi$ phase shift\\
        \hline
        $\nu=\frac{3}{3M+2}$ fermionic Read-Rezayi ($M$ odd)& $\text{Fib}\times\Z_{2(3M+2)}$& universal neutral-channel amplitude\\
        $\nu=\frac{3}{3M+2}$ bosonic Read-Rezayi ($M$ even)& $\text{Fib}\times\Z_{(3M+2)/2}$& ratios and a relative $\pi$ phase shift\\
        \hline
    \end{tabular}
    \caption{Summary of the zero mode structures and characteristic tunneling signatures of FQH-SCs discussed in Secs.~\ref{sec:1/m} and \ref{sec:nonab} and Appendix~\ref{app:abelian}. The neutral factor $\Z_2^{(n)}$ in the generalized Moore–Read states originates from the Ising fusion channels, while the Read–Rezayi state supports a non-Abelian Fibonacci neutral sector. For the RR FQH-SC, the magnitude ratio of the two neutral-channel amplitudes is $\Phi^{-2}$ for tunneling of zero mode operators and $\Phi^{-1}$ for direct zero mode anyon transfer, where $\Phi=(1+\sqrt{5})/2$ is the golden ratio.}
    \label{tab:zms}
\end{table}

Applying this framework, we revisit the zero mode structures and fusion-channel-dependent tunneling signatures of Laughlin states \cite{clarke-2013,lindner-2012,cheng-2012}, which we extend to generalized Moore–Read (MR) and $k=3$ Read–Rezayi (RR) FQH-SCs. For fermionic Laughlin FQH-SCs at filling $\nu=1/m$, we recover the familiar $\Z_{2m}$ parafermion zero mode structure. For generalized MR states at filling fraction $\nu=1/m$, we obtain a $\Z_2^{(n)}\times\Z_{N}$ zero mode structure, with $N=2m$ for fermionic states and $N=m$ for bosonic states. Zero mode tunneling produces a Josephson spectrum that depends on both the charge and neutral defect fusion channels. The two neutral fusion channels have equal tunneling magnitudes but differ by a relative $\pi$ phase shift. For the $k=3$ RR state at filling $\nu=3/(3M+2)$, we find a $\text{Fib}\times\Z_{N}$ zero mode structure, where $\text{Fib}\equiv\{1,\tau\}$ stands for the Fibonacci topological order \cite{bonderson-2007-thesis} and $N=2(3M+2)$ for fermionic states and $N=(3M+2)/2$ for bosonic states. In this case, tunneling of zero mode operators and direct zero mode anyon transfer yield different universal neutral-sector amplitude ratios. This difference provides a characteristic signature of the underlying Fibonacci defect structure. Table~\ref{tab:zms} summarizes the zero mode structures and tunneling signatures considered in this work. 

We want to remind the readers that defect braiding alone does not provide a universal set of gates for quantum computation in the Abelian or MR examples considered here as neither $\Z_N$ PZMs \cite{hastings-2013,hastings-2014,hutter-2016} nor the additional Ising neutral sector in the MR FQH-SC support universal TQC. Among the examples studied here, only the RR FQH-SC supports defects carrying a Fibonacci neutral sector, whose braiding enables universal TQC \cite{freedman-2000}. For the RR FQH-SC, the superconducting geometry also defines the defect positions and allows their couplings to be tuned. This offers a route to controlled braiding gates through coupling sequences or topological-charge measurements \cite{bonderson-2008,bonderson-2009,bonderson-2013}, without requiring the physical transport of bulk Fibonacci anyons in the RR FQH state.

Recent developments in two-dimensional van der Waals materials strengthen the prospects for realizing FQH-SCs while broadening the range of candidate parent topological states to include non-Abelian phases. Systems such as graphene and transition-metal dichalcogenides can host both topological order and superconductivity, either along proximity-coupled edges \cite{lee-2017,shi-2020,gul-2022,jia-2024} or in adjacent regions of a common phase diagram \cite{han-2025,xu-2025-1}. Fractional quantum anomalous Hall states (FQAH) have also been observed in the absence of an external magnetic field \cite{neupert-2011,sheng-2011,regnault-2011,cai-2023,zeng-2023,park-2023,lu-2024}, potentially enhancing their compatibility with superconductivity. In parallel, theoretical and numerical studies have identified candidate non-Abelian phases in moir\'{e} and multilayer systems \cite{kang-2024,zhang-2024,reddy-2024,ahn-2024,chen-2025,abouelkomsan-2025,may-mann-2025,jian-2025,liu-2025}.

In a recent work \cite{cao-2026-l}, we proposed Josephson and Fabry–P\'{e}rot configurations for probing superconducting defects in FQH states. That work studied stochastic switching between branches of zero mode tunneling spectra via anyon tunneling and interferometric signatures in the weak and strong zero mode tunneling regimes, including both Laughlin and Moore–Read parent states. Here, we develop the underlying bulk-to-boundary construction of localized zero mode operators and distinguish their tunneling from direct zero mode anyon transfer. We also extend this construction to Read–Rezayi FQH-SCs, whose Fibonacci neutral sector produces distinct fusion-channel-dependent tunneling amplitudes.

This paper is organized as follows. In Sec.~\ref{sec:setup}, we review the charge-conjugation defect description of FQH-SCs, construct localized zero mode operators from bulk Wilson operators, and describe two tunneling processes and their fusion-channel dependence. In Sec.~\ref{sec:1/m}, we apply this framework to Laughlin FQH-SCs and recover their parafermion algebra and fractional Josephson response. Section \ref{sec:nonab} extends this construction to FQH-SCs with generalized Moore–Read and Read-Rezayi parent states, compares candidate phases at $\nu=1/2$, and derives their neutral-sector tunneling signatures. Technical details concerning defect theory, fermionic modular extensions, renormalization-group (RG) analysis, and additional Abelian examples are presented in the appendices.

\section{General Framework: FQH-SC defects and edge zero mode operators}\label{sec:setup}
\subsection{Physical setup and edge theory}
\begin{figure}
    \centering
    \includegraphics[scale=1]{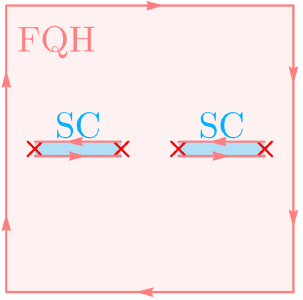}
    \caption{Counter-propagating FQH edges (pink) proximity-coupled by superconductors (cyan), with zero modes localized at the endpoint defects of each FQH-SC segment.}
    \label{fig:setup}
\end{figure}

We consider the FQH-SC geometry of Refs.~\cite{barkeshli-2013-genon,cong-2016,cong-2017,calzona-2023}, in which two long, narrow superconducting islands are embedded in a fractionalized state $\C$ , as shown in Fig. \ref{fig:setup}. Such a structure may be realized by forming trenches in the bulk state and proximity coupling their counter-propagating edges to superconductors \cite{lee-2017,gul-2022} or by inducing superconductivity through an on-chip metallization process \cite{jia-2024}. Although the construction applies equally to FQH, FQAH, and related fractionalized phases, we refer to the parent phase as an FQH state for brevity. We refer to each resulting gapped, superconductor-proximitized interface as an FQH-SC segment.

The theory of FQH-SC consists of counter-propagating edge modes coupled by an SC proximity effect. The edges are described by chiral and anti-chiral conformal field theories (CFTs) inherited from the parent FQH state and the proximity effect takes the form of a Cooper-pairing perturbation \cite{clarke-2013,lindner-2012,cheng-2012} 
\begin{align}
    \L_{\text{FQH-SC}}=\L_{CFT,+}+\L_{CFT,-}+\pbr{\De \Psi_+\Psi_-+\hc}+...\label{eqn:fqhsc}\,.
\end{align}
The Lagrangians $\L_{CFT,\pm}$ in \eqref{eqn:fqhsc} describe the chiral and anti-chiral edges associated with the FQH state $\C_+\equiv \C$ and its parity-reversed copy $\C_-$, respectively. The operators $\Psi_\pm$ are the corresponding local physical particle operators (electrons for fermionic parent states), and $\De$ is the superconducting order parameter. The ellipsis denotes additional local interactions, including density-density interactions that modify the scaling dimension of the Cooper-pairing perturbation. To induce pairing in a spin-polarized FQH state, the SC either needs to be of spin-triplet nature or to exhibit Rashba spin-orbit coupling that serves as a spin-flip mechanism \cite{gul-2022}. Throughout this work, we assume that Cooper pairing fully gaps the counter-propagating edges, either because the pairing perturbation is relevant and flows to strong coupling or because its bare amplitude $\De$ is sufficiently large, with no additional modes generated by edge reconstruction \cite{chamon-1994,aleiner-1994,kane-1994,wan-2002}. The conditions for the pairing perturbation to become relevant depend on the parent edge theory and the interactions between the counter-propagating modes (see details in App.~\ref{app:rg}).

An alternative gapping mechanism using charge-conserving electron tunneling between counter-propagating edges in bilayer FQH systems was proposed in Ref. \cite{barkeshli-2014}. The resulting line junctions exchange the layer labels of traversing anyons, whereas the superconducting interfaces considered here implement charge conjugation, as discussed in the following subsection. The corresponding interferometric proposals probe the defect fusion space through current-noise cross-correlations in bilayer FQH systems \cite{barkeshli-2014}, and through fusion-channel switching and suppression of interference by strong zero mode tunneling in FQH-SCs \cite{cao-2026-l}.

\subsection{FQH-SC as charge-conjugation defect line}
\label{subsec:defect}

\begin{figure}
    \centering
    \subfloat[]{%
        \includegraphics[width=0.3\columnwidth]{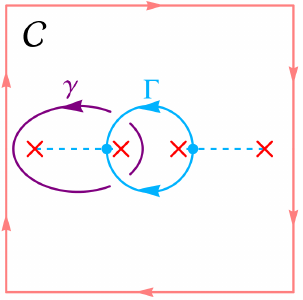}
        \label{fig:setup1}
    }\qquad
    \subfloat[]{%
        \includegraphics[width=0.3\columnwidth]{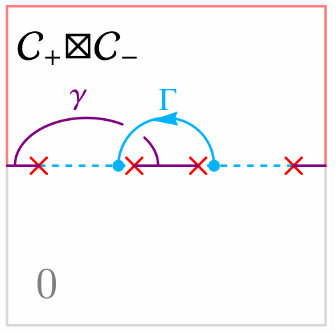}
        \label{fig:setup_fold}
    }
    \caption{\textbf{(a)} The two classes of Wilson operators considered in this work. Directed lines represent chiral edges or oriented Wilson lines, whereas undirected lines represent nonchiral edges or unoriented Wilson lines. The Wilson loop $W(\ga)$ (purple) acts within the ground-state subspace, while the Wilson line $W(\Ga)$ (cyan) terminates at the superconducting segments and represents zero mode tunneling. \textbf{(b)} Folded representation of panel (a), in which both Wilson operators terminate on gapped boundaries of the folded theory.}
    \label{fig:setup_and_fold}
\end{figure}
Although the one-dimensional bulk of a finite FQH-SC segment is fully gapped in the superconducting phase, its two endpoints host localized zero modes \cite{kitaev-2001,clarke-2013,cheng-2012,lindner-2012}. These endpoints are extrinsic defects with well-defined topological charges and fusion rules \cite{barkeshli-2013-genon,barkeshli-2019}. When several endpoint defects are present, their allowed fusion channels, subject to any constraint on the total topological charge, span a topologically degenerate low-energy subspace. Local perturbations near an isolated endpoint cannot distinguish these channels, whereas coupling between nearby endpoints can split the degeneracy in a fusion-channel-dependent manner \cite{bonderson-2009-split}.  The non-Abelian structure therefore arises from the collective fusion space of several extrinsic defects, rather than from an isolated zero mode operator.

An equivalent boundary description is obtained by folding the two-dimensional system along the superconducting segment \cite{barkeshli-2013,barkeshli-2013-genon}, as illustrated in Fig. \ref{fig:setup_and_fold}. The folded system contains two copies of the parent topological order with opposite orientations, denoted by $\C_+\boxtimes\C_-$, where $\boxtimes$ denotes stacking of TOs. Excitations in the folded system, denoted by $(a,b)$, carry one anyon label from each copy and correspond to the fusion product $a\times b$ in the unfolded theory. The superconducting interface is then described as a gapped boundary \cite{kitaev-2012,kong-2014,lan-2015,hung-2015,cong-2016,cong-2017,cong-2017-2,lan-2020} of $\C_+\boxtimes\C_-$, where diagonal anyons $(a,a)$ condense. 

\begin{figure}[ht!]
     \centering
     \includegraphics[scale=1]{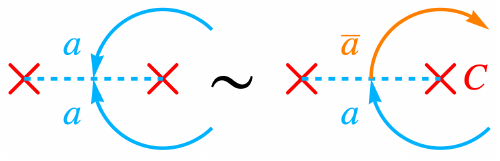}
     \caption{A superconducting segment that condenses the diagonal anyons $(a,a)$ acts as a charge-conjugation defect line. An anyon $a$ crossing the segment is transformed into its conjugate $\bar{a}$, and the two endpoints of the segment are charge-conjugation defects.}
     \label{fig:SC=defect}
\end{figure}

Note that in the FQH-SC setup, the condensation of anyons occurs at the boundary: a Wilson line operator $W_a(\Ga)$ acquires a vacuum expectation value in the ground state of FQH-SC. 
This should be distinguished from bulk ``anyon condensation" \cite{bais-2009,bais-2009-2,burnell-2018}, i.e. formally gauging an anomaly-free one-form symmetry \cite{cheng-2026}. 
In the ``anyon condensation" picture, anyons with nontrivial monodromy with a condensed anyon are confined, and anyons related by fusion with condensed anyons are identified. 
By contrast, condensation at the FQH-SC boundary leaves the bulk topological order unchanged, so the corresponding noncontractible Wilson loops remain as symmetry operators acting on the defect Hilbert space. In this sense, each condensed diagonal anyon $(a,a)$ may be viewed as an ``anyon Cooper pair", formed from one $a$ anyon on each counter-propagating edge. 
Equivalently, in the unfolded picture, an anyon line crossing the superconducting segment is transformed according to $\rho_{\vb{C}}(a)=\bar{a}$, as illustrated in Fig.~\ref{fig:SC=defect}. 
For the examples considered in this work, electric charge conjugation, $q\mapsto -q$, induces the same action on the anyon sectors as topological charge conjugation, 
$a\mapsto\bar{a}$. We therefore refer to both as charge conjugation. 
The superconducting segment therefore acts as a charge-conjugation branch cut, and its two endpoints are extrinsic charge-conjugation defects.

The $G$-crossed braided tensor category is a mathematical formalism that extends the parent bosonic topological order to include symmetry defects and their fusion and braiding data \cite{etingof-2010,barkeshli-2019}. Here $G$ is a symmetry group of the parent topological order, acting on its anyons in a way that preserves their fusion and braiding data. Each element $\vb{g}\in G$ specifies the symmetry action across a defect branch cut, and the defect sector $\C_{\vb{g}}$ contains the corresponding defect types (see details in App.~\ref{subapp:Gcross}). For charge conjugation,  $G=\Z_2^C\equiv\cbr{\vb{1},\vb{C}}$, so the extended theory, called a $G$-graded fusion category \cite{barkeshli-2019}, is
\begin{align}
    \C_G=\C_{\vb{1}}\oplus\C_{\vb{C}}\,,
\end{align}
where $\C_{\vb{1}}\equiv\C$ is the parent FQH state topological order, and $\C_{\vb{C}}$ contains the charge conjugation $\vb{C}$-defects. Distinct topological charges of an endpoint within the same symmetry sector are called defect types. Attaching an ordinary anyon can change the defect type without changing the branch-cut action. The sector label $\vb{C}$ therefore does not uniquely specify the endpoint defect (see details in App.~\ref{subapp:Gcross}).

We denote the two endpoint defects by $X_{\vb{C}}$, $\bar{X}_{\vb{C}}\in\C_{\vb{C}}$ chosen to be conjugates so that their fusion contains the vacuum. Because $\vb{C}^2=\vb{1}$, their fusion lies in the anyon sector $\C_{\vb{1}}$:
\begin{align}
    X_{\vb{C}}\times \bar{X}_{\vb{C}}=\sum_{c\in\C_{\vb{1}}} N_{X_{\vb{C}}\bar{X}_{\vb{C}}}^c\,c\,,
\end{align}
where $N_{X_{\vb{C}}\bar{X}_{\vb{C}}}^c$ is the fusion multiplicity, the number of independent fusion states with total charge $c$. The sum includes the vacuum $c=0$, with $N^0_{X_{\vb{C}}\bar{X}_{\vb{C}}}=1$ for a conjugate pair of simple defects. Here $c$ labels the fusion outcome of the pair, whereas $X_{\vb{C}}$ and $\bar{X}_{\vb{C}}$ specify the individual endpoint defect types (see details in App.~\ref{subapp:defectf}). The allowed fusion channels label the topological charge stored nonlocally between the defects and span the defect Hilbert space. 

Within this framework, an anyon $c_{\text{zm}}\in\C_{\vb{1}}$ is a zero mode anyon of $X_{\vb{C}}$ if it can be absorbed or emitted by an endpoint defect without changing its defect type \cite{barkeshli-2019}. In terms of fusion rules, this means that $X_{\vb{C}}$ is an allowed fusion outcome of $X_{\vb{C}}$ and $c_{\rm zm}$:
\begin{align}
    X_{\vb{C}}\times c_{\text{zm}}\ni X_{\vb{C}}\,,\quad N_{X_{\vb{C}}c_{\text{zm}}}^{X_{\vb{C}}}> 0\,.\label{eqn:Gzm_def}
\end{align}
Whether $c_{\rm zm}$ is a zero mode anyon therefore depends on the defect type: an anyon absorbed by one type need not be absorbed by another (see details in App.~\ref{subapp:defectf}).

\subsection{Criterion for zero mode anyons}
\begin{figure}
    \centering
    \includegraphics[scale=1]{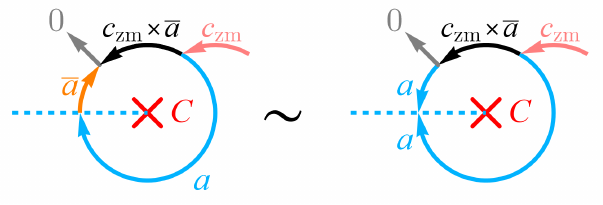}
    \caption{Bulk fusion criterion for a candidate zero mode anyon $c_{\text{zm}}$ of an FQH-SC endpoint. The anyon $a$ crosses the charge-conjugation defect line and becomes $\bar{a}$. The process closes without leaving a bulk excitation when $c_{\text{zm}}\in a\times a$.}
    \label{fig:zm_def}
\end{figure}

To connect zero mode anyons with their localized operator realizations, we derive a criterion that identifies candidate zero mode anyons from the fusion rules of the parent FQH state and the charge-conjugation action. Because the diagonal anyons $(a,a)$ condense on the folded superconducting boundary, the corresponding Wilson lines can terminate there without creating a topologically nontrivial boundary excitation. The termination defines a localized operator labeled by $a$, while the relation between $a$ and $c_{\rm zm}$ is determined by the unfolded process shown in Fig.~\ref{fig:zm_def}. In this process, the zero mode anyon $c_{\text{zm}}$ first splits into an anyon $a$ and an allowed fusion outcome of $c_{\text{zm}}\times\bar{a}$. The anyon $a$ then crosses the charge-conjugation defect line and transforms into $\bar{a}$. The process leaves no additional bulk excitation when the selected fusion outcome can fuse with the transformed anyon $\bar{a}$ into the vacuum, giving the condition \footnote{For a $\vb{g}$-defect, where $\vb{g}\in G$ labels the symmetry label carried by the defect, this condition becomes $c_{\rm zm}\times\bar{a}\times\rho_{\vb{g}}(a)\ni 0$ or equivalently $c_{\rm zm}\in a\times \rho_{\vb{g}}(\bar{a})$.}
\begin{align}
    c_{\text{zm}}\times\bar{a}\times\bar{a}\ni0\,,\quad N_{aa}^{c_{\text{zm}}}>0 \label{eqn:zm_def}
\end{align}
where $0$ denotes the vacuum sector. Equivalently, $c_{\text{zm}}\in a\times a$. For Abelian anyons, this condition reduces to $c_{\text{zm}}= a\times a$. 

Equation \eqref{eqn:zm_def} determines candidate zero mode anyons using only the fusion rules of the parent FQH topological order and the charge-conjugation action. For fermionic parent states, the Wilson-line label $a$ may lie in the minimal modular extension introduced in App.~\ref{subapp:mme}, while $c_{\rm zm}$ is restricted to the physical anyon sector. Because this criterion does not specify the endpoint defect types, each candidate must additionally satisfy the condition in Eq.~\eqref{eqn:Gzm_def}. A zero mode anyon $c_{\rm zm}$ labels a topological charge that the defect can absorb, whereas a zero mode operator is associated with the termination of a Wilson line labeled by $a$. Their correspondence need not be one-to-one: different Wilson-line labels $a$ can realize the same zero mode anyon, while a non-Abelian fusion product $a\times a$ can contain several zero mode anyons supported by the chosen defect. We use the relation in Eq. \eqref{eqn:zm_def} to construct localized zero mode operators acting on the defect fusion space in the effectively one-dimensional FQH-SC theory.

The $\nu=1$ integer quantum Hall-SC provides a simple example as its endpoints host Majorana zero modes \cite{kitaev-2001,clarke-2013,cheng-2012,lindner-2012}. There are no nontrivial anyons in the theory, so the parent theory is described by $\{0,\Psi\}$ where $0$ denotes the vacuum and $\Psi$ denotes the physical electron. The endpoint defects have the Ising-type fusion structure familiar from vortices in a $p+ip$ superconductor in the weak-pairing limit \cite{read-2000,ivanov-2001}. A conjugate pair of endpoint defects obeys $X_{\vb{C}}\times\bar{X}_{\vb{C}}=0+\Psi$. The two fusion channels correspond to the unoccupied and occupied states of the fermionic mode formed from the endpoint operators $\ga_1,~\ga_2$. Because $X_{\vb{C}}\times\Psi=X_{\vb{C}}$, the physical fermion $\Psi$ is a zero mode anyon according to Eq.~\eqref{eqn:Gzm_def}, whereas $\ga_i$ is its localized operator realization. Here the non-Abelian fusion structure belongs to the defects: $\Psi$ is a fermion, and $\ga_i$ a Majorana fermionic operator acting on the collective defect fusion space, not an Ising anyon. Coupling the endpoints through $i\ga_1\ga_2$ distinguishes the two fusion channels and probes the charge stored nonlocally between the defects. These Majorana zero modes are the $m=1$ case of the $\Z_{2m}$ parafermion zero modes reviewed in Sec.~\ref{sec:1/m}.

\subsection{Wilson operators and edge zero mode operators}
\label{subsec:Wilson}

The Wilson-line construction of localized zero mode operators was established for Abelian parent topological orders in Ref. \cite{barkeshli-2013}. Here we extend this construction to non-Abelian parent states using the two-segment geometry shown in Fig.~\ref{fig:setup1}. This geometry supports two natural classes of Wilson operators. The first is a Wilson line $W_a(\Ga)$, which transfers anyonic charge between the two FQH-SC segments and carries the label $(a,a)$ in the folded description. Because $(a,a)$ is condensed on both superconducting boundaries, $W_a(\Ga)$ acts within the ground-state subspace and provides the bulk representation of zero mode tunneling between the segments. The second operator is a noncontractible Wilson loop $W_b(\ga)$, which transports an anyon $b$ around one FQH-SC segment. Such paths can be implemented through quasiparticle propagation along the edges and tunneling across quantum point contacts (QPCs) (see Ref. \cite{cao-2026-l}). In a Fabry–P\'{e}rot interferometer \cite{chamon-1997}, interference between two tunneling paths probes the topological charge enclosed by the resulting loop \cite{fradkin-1998,bonderson-2006,stern-2006}. Their algebra therefore describes how the parent FQH anyon data act on the nonlocal ground-state degrees of freedom. The Wilson loop operator $W_b(\ga)$ acts on the anyon line $a$ with an amplitude determined by the loop-removing algebra in Refs.~\cite{verlinde-1988,witten-1989,kitaev-2006}:
\begin{align}
    W_b(\ga)W_a(\Ga)=\pbr{\frac{\S_{ab}}{\S_{a0}}}^2W_{a}(\Ga)=
    \pspicture[shift=-0.55](-0.25,-0.1)(0.9,1.3)
\small
  \psset{linewidth=0.9pt,linecolor=black,arrowscale=1.5,arrowinset=0.15}
  \psline(0.4,0)(0.4,0.22)
  \psline(0.4,0.45)(0.4,1.2)
  \psline(0.6,0)(0.6,0.22)
  \psline(0.6,0.45)(0.6,1.2)
  \psellipse[linewidth=0.9pt,linecolor=black,border=0](0.5,0.55)(0.5,0.23)
  \psset{linewidth=0.9pt,linecolor=black,arrowscale=1.4,arrowinset=0.15}
  \psline{-<}(0.2,0.37)(0.3,0.34)
\psline[linewidth=0.9pt,linecolor=black,border=2.5pt,arrows=->,arrowscale=1.5,
arrowinset=0.15](0.4,0.5)(0.4,1.0)
\psline[linewidth=0.9pt,linecolor=black,border=2.5pt,arrows=->,arrowscale=1.5,
arrowinset=0.15](0.6,0.5)(0.6,1.0)
  \rput[bl]{0}(-0.15,0.15){\ensuremath{b}}
  \rput[tr]{0}(0.33,1.2){\ensuremath{a}}
  \rput[tl]{0}(0.65,1.2){\ensuremath{a}}
  \endpspicture
\,\,=\pbr{\frac{\S_{ab}}{\S_{a0}}}^2
\pspicture[shift=-0.55](0.05,-0.1)(1,1.3)
\small
  \psset{linewidth=0.9pt,linecolor=black,arrowscale=1.5,arrowinset=0.15}
  \psline(0.4,0)(0.4,1.2)
\psline[linewidth=0.9pt,linecolor=black,arrows=->,arrowscale=1.5,
arrowinset=0.15](0.6,0.5)(0.6,0.75)
  \psline(0.6,0)(0.6,1.2)
\psline[linewidth=0.9pt,linecolor=black,arrows=->,arrowscale=1.5,
arrowinset=0.15](0.4,0.5)(0.4,0.75)
  \rput[tl]{0}(0.65,1.2){\ensuremath{a}}
  \rput[tr]{0}(0.33,1.2){\ensuremath{a}}
  \endpspicture\,,\label{eqn:wilson}
\end{align}
where $\S_{ab}$ denotes an element of the modular $\S$-matrix, which encodes the mutual braiding data of the anyons $a$ and $b$, and $0$ denotes the vacuum sector. For a bosonic parent topological order, $\S$ is directly evaluated in $\C$, whereas for a fermionic parent topological order, it is evaluated in a minimal modular extension (see details in App.~\ref{subapp:mme}). The dependence of the loop amplitude on $a$ determines which topological charge sectors can be distinguished by the Wilson loop.

Folding the geometry in Fig.~\ref{fig:setup1} maps the bulk Wilson operators to the Wilson arcs terminating at different boundaries, as shown in Fig.~\ref{fig:setup_fold}. Specifically, $W_a(\Ga)$ maps to an arc labeled by $(a,a)$, connecting the superconducting boundaries, whereas $W_b(\ga)$ maps to an arc labeled by $(b,\bar{b})$ terminating on the trivial folded boundary. The folded Wilson arcs obey 
\footnote{In gapped edges of general nonchiral topological order, the algebra between condensed and confined anyon lines is determined by half-linking \cite{kapustin-2011,shen-2019}, which depends on the correspondence between bulk anyons and anyons in the boundary spectrum. For the particular cases of FQH-SC, since the bulk topological order is a folded theory, the half-linking always coincides with the bulk $\S$-matrix \cite{shen-2019}. We thank Da-Chuan Lu for pointing us to Ref. \cite{shen-2019}.}
\begin{align}
    W_{(b,\overline{b})}W_{(a,a)}=\frac{\S^\C_{ab}}{\S^\C_{a0}}\frac{\S_{a\overline{b}}^{\bar{\C}}}{\S_{a0}^{\bar{\C}}}W_{(a,a)}=\pbr{\frac{\S_{ab}}{\S_{a0}}}^2W_{(a,a)}\,.\label{eqn:fold_wilson}
\end{align}
The $\S$-matrix of the parity-reversed theory $\bar{\C}$ is the complex conjugate of that of $\C$. Together with $\S_{a\overline{b}}=\S_{ab}^\ast$, this gives $\S_{a\overline{b}}^{\bar{\C}}=\S_{ab}^\C$, yielding the squared factor in Eq.~\eqref{eqn:fold_wilson}. In the language of boundary anyon condensation, $a$ labels condensed anyons $(a,a)$ on the superconducting boundary, while noncondensed sectors $(b,\bar b)$ with nontrivial monodromy with the condensed anyons cannot terminate there without leaving a boundary excitation \cite{kitaev-2012,kong-2014}.

The Cooper-pairing term modifies the boundary dynamics but preserves the equal-time exchange relations inherited from the bulk Wilson algebra. For Wilson operators that preserve the ground-state subspace, these relations therefore also hold after projection onto the defect fusion space. Each endpoint of the folded Wilson arc $W_{(a,a)}(\Ga)$ is represented by a localized zero mode operator $\al_a^{(i)}$, where $i=1,2$ labels the two segments. Consequently, the complete Wilson arc connecting two segments projects to a zero mode tunneling operator, while the Wilson loop projects to an operator $U_b$ acting within the defect Hilbert space:
\begin{align}
    W_{(a,a)}\longrightarrow\al_a^{(1)\da}\al_a^{(2)}\,,\quad W_{(b,\overline{b})}\longrightarrow U_b\,.
\end{align}
Here the arrows denote equivalence after projection onto the defect Hilbert space. We henceforth suppress the folded-pair labels and retain only the parent anyon labels $a$ and $b$. We refer to $\al_a$ and $U_b$ in the one-dimensional theory as zero mode operators and projected loop operators, respectively.

In the parent FQH edge CFTs, the endpoints of bulk Wilson arcs correspond to primary field insertions \cite{witten-1989,moore-1989,wen-1990,wen-1991,wen-1995,lopez-1999}. In the gapped FQH-SC theory, we express the zero mode operators and projected loop operators as products of these fields:
\begin{align}
    \al_a(t,x)\sim O_{a_+}(t,x) O_{a_-}(t,x)\,,\quad U_b(t,x)\sim O_{b_+}(t,x) O_{\overline{b}_-}(t,x)\,,\label{eqn:edge_rep}
\end{align}
where $O_{a_+}$ and $O_{a_-}$ are the primary field operators associated with anyon $a$ in the chiral and anti-chiral edge theories, respectively. Likewise, $O_{b_+}$ and $O_{\bar{b}_-}$ are associated with $b$ and $\bar{b}$, respectively, consistent with the folded label $(b,\bar{b})$. Here, $\sim$ denotes low-energy equivalence after projection onto the defect fusion space, with nonuniversal normalization and short-distance details suppressed.

The local edge representation in Eq.~\eqref{eqn:edge_rep} allows us to formulate the Wilson-loop action on individual zero mode operators. Although $W_{(a,a)}$ projects onto the tunneling operator $\al_a^{(1)\da}\al_a^{(2)}$, the action on a single factor $\al_a$ is evaluated by linking the folded line $(b,\bar{b})$ with the corresponding endpoint insertion $(a,a)$. Because the projected operator $U_b$ need not be invertible, we denote its action on zero mode operators by $\U_b[\cdot]$. The projected loop-removal relation is then
\begin{align}
    \U_b[\al_a]=\pbr{\frac{\S_{ab}}{\S_{a0}}}^2\al_a\,. \label{eqn:zmalg}
\end{align}
For an invertible line represented by a unitary operator $U_b$, this action reduces to ordinary conjugation,
\begin{align}
     \U_b[\al_a]= U^\da_{b}\al_aU_b\,.
\end{align}
For non-invertible $U_b$, $\U_b[\cdot]$ instead denotes the corresponding topological-line action \cite{chang-2019} and should not be interpreted as conjugation by an inverse symmetry operator. Equation \eqref{eqn:zmalg} determines the action of $U_b$ on the zero mode operator $\al_a$ from the same modular data that enter the bulk Wilson algebra. For Abelian FQH-SCs, Eq.~\eqref{eqn:zmalg} reproduces the familiar symmetry action on PZMs \cite{fendley-2012}. For non-Abelian FQH-SCs, it specifies the Wilson-line action on zero mode operators within the fusion space of the charge-conjugation defects.

The bulk-to-boundary projection converts bulk linking relations into spatially ordered operator relations. For an FQH-SC segment occupying $x\in[0,L]$, we place the projected loop operators at $x=0,L$, and the zero mode operators slightly inside the segment, at $x=\vep$ and $x= L-\vep$, where $\vep$ is infinitesimal. This convention implements the action in Eq.~\eqref{eqn:zmalg}. 

A projected loop operator represents a symmetry only when its action is compatible with the boundary Hamiltonian. For invertible $U_b$, one must therefore verify $\comm{H_{\text{FQH-SC}}}{U_b}=0$ in the effective one-dimensional FQH-SC theory. For non-invertible $U_b$, the corresponding line action must remain topological: it must be deformable through insertions of the pairing perturbation without changing correlation functions. Under this condition, its fusion and crossing relations survive the (RG) flow to the gapped theory \cite{chang-2019}.

For electronic FQH-SCs---where microscopic degrees of freedom are fermions---the ground-state degeneracy may contain an additional contribution associated with fermion parity \cite{kitaev-2001,fidkowski-2010,fidkowski-2011,turner-2011,barkeshli-2013}. We incorporate this contribution by embedding the fermionic topological order into a minimal modular extension \cite{bruillard-2017,aasen-2022,kobayashi-2022}, an auxiliary modular theory containing the physical anyons together with fermion-parity flux sectors (see details in App.~\ref{subapp:mme}) \footnote{We follow the convention common in the condensed matter and CFT literature: physical anyon sectors are local with respect to the electron $\Psi$, and sectors related by fusion with $\Psi$ are identified. In the super-modular description of fermionic topological order \cite{bonderson-2007-thesis,bruillard-2017}, the electron is instead retained as a distinct transparent ``anyon''. The resulting category has transparent subcategory $\text{sVec}\equiv\{0,\Psi\}$: the vacuum and the electron are the only simple objects with trivial monodromy with every anyon. The minimal modular extension used in this work is the same in both conventions.}. The Wilson algebra and its bulk-to-boundary projection are then evaluated in the extended modular theory, while zero mode anyons are retained in the FQH-SC theory. This construction resolves fermion-parity information that braiding among the physical anyons alone cannot distinguish.

\subsection{Zero mode tunneling and fusion-channel-dependent signatures}\label{subsec:tunnel}
When two defects are brought into proximity, they may couple through two physically distinct processes, thereby lifting the topological ground-state degeneracy \cite{bonderson-2009-split}. The first is zero mode tunneling $\al_a^{(1)\da}\al_a^{(2)}$, whereas the second is direct transfer of a zero mode anyon $c_{\text{zm}}$ between the defects. These processes need not have the same fusion-channel dependence.

Zero mode operator tunneling is governed by an effective Hamiltonian whose matrix elements in the defect fusion space determine the fusion-channel-dependent spectrum. At energies below both the FQH quasiparticle gap and the superconducting gap of the proximitized edges, this Hamiltonian takes the form \cite{clarke-2013,cheng-2012,lindner-2012},
\begin{align}
    H_{\text{tunneling}}=\sum_a \Ga_a(\phi_{\text{SC}}) \al_a^{(1)\da}\al_a^{(2)}+\hc\,,\label{eqn:tunnel}
\end{align}
where $\Ga_a(\phi_{\text{SC}})$ is a nonuniversal tunneling amplitude whose phase depends on the superconducting phase difference $\phi_{\text{SC}}$ and on the electric charge transferred by the tunneling process. Determining the fusion-channel-dependent spectrum therefore amounts to evaluating the matrix elements of $\al_a^{(1)\da}\al_a^{(2)}$ in the defect Hilbert space. For Abelian FQH-SCs, the tunneling matrix elements can be evaluated explicitly using the bosonic vertex operator expressions for the PZMs \cite{clarke-2013,cheng-2012,lindner-2012}, allowing $H_{\text{tunneling}}$ to be diagonalized explicitly. For generic non-Abelian FQH-SCs, however, the primary field matrix elements cannot generally be reduced to free-field operators. We therefore determine the fusion-channel dependence of the tunneling spectrum from the corresponding bulk Wilson processes rather than by directly evaluating the zero mode operator matrix elements.
\begin{figure}[ht]
    \centering
    \subfloat[]{%
        \includegraphics[width=0.6\columnwidth]{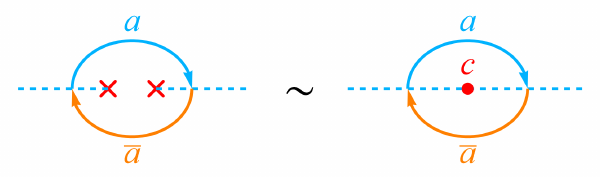}
        \label{fig:zm_tunnel1}
    }
    \subfloat[]{%
        \includegraphics[width=0.35\columnwidth]{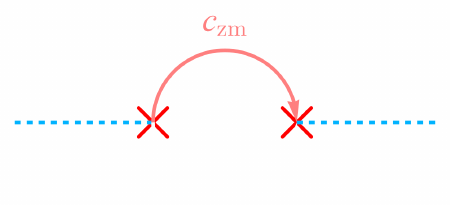}
        \label{fig:zm_tunnel2}
    }
    \hfill
    \caption{\textbf{(a)} Wilson-loop measurement of the fusion channel $c$ of two charge-conjugation defects. A loop labeled by $a$ encloses the defects and crosses the two defect lines, giving the monodromy $M_{ac}$. \textbf{(b)} Direct transfer of the zero mode anyon $c_{\text{zm}}$ between $X_{\vb{C}}$ and $\bar{X}_{\vb{C}}$. Reassociating the corresponding fusion tree gives the $G$-crossed $F$-matrix element $\sbr{F_{c}^{X_{\vb{C}}c_{\text{zm}}\bar{X}_{\vb{C}}}}_{X_{\vb{C}}\bar{X}_{\vb{C}}}$.}
    \label{fig:zm_tunnel}
\end{figure}

The bulk Wilson process associated with zero mode operator tunneling reduces to a monodromy loop around the two endpoint defects. This procedure works as follows. We denote their fusion channel by $c$, which should not be confused with the zero mode anyon label $c_{\text{zm}}$. Under the bulk-to-boundary projection, the zero mode tunneling operator $\al_a^{(1)\da}\al_a^{(2)}$ is represented by the folded Wilson arc $W_{(a,a)}(\Ga)$. In the unfolded defect geometry, this arc may be deformed into a closed Wilson loop labeled by $a$ that surrounds the two endpoint defects and crosses their charge-conjugation defect lines twice, as shown in Fig.~\ref{fig:zm_tunnel1}. For the multiplicity-free defect fusion spaces considered here (see details in App.~\ref{subapp:defectf}), this loop acts diagonally in the defect fusion-channel basis \cite{verlinde-1988,witten-1989,kitaev-2006}: 
\begin{align}
    W_a\ket{c}=\frac{\S_{ac}\S_{00}}{\S_{a0}\S_{c0}}\ket{c}=M_{ac}\ket{c}\,,\label{eqn:probe}
\end{align}
where $M_{ac}$ is the normalized monodromy scalar associated with taking $a$ around the total topological charge $c$ of the defect pair. The normalization is chosen such that the Wilson loop acts as the identity in the vacuum fusion channel: $W_{a}\ket{0}=\ket{0}$.

With this normalization, the fusion-channel-dependent spectrum is
\begin{align}
    E_c(\phi_{\text{SC}})=\sum_a \sbr{\Ga_a(\phi_{\text{SC}})M_{ac}+\Ga_a(\phi_{\text{SC}})^\ast M_{ac}^\ast}\,.\label{eqn:JJ}
\end{align}
Any factor independent of the fusion channel that relates the microscopic tunneling operator to the normalized Wilson loop is absorbed into the amplitude $\Ga_a$, where $\phi_{\rm SC}$ is the phase difference between the two superconductors. The folded Wilson line $W_{(a,a)}$ transfers charge $2q_a$, where $q_a$ is the electric charge of anyon $a$. This transferred charge fixes the superconducting-phase dependence of $\Ga_a$. For example, if a tunneling process transfers charge $2e/\ka$, then $\Ga_a(\phi_{\text{SC}})\propto e^{-i\phi_{\text{SC}}/\ka}$ and its contribution to a fixed fusion-channel branch is $2\pi \ka$-periodic. This enlarged periodicity is the fractional Josephson effect \cite{kitaev-2001,clarke-2013,cheng-2012,lindner-2012}. Because $M_{ac}$ is determined by the modular data of the parent FQH state or its chosen minimal modular extension, Eq.~\eqref{eqn:JJ} applies even when the corresponding zero mode operators do not admit a free-field representation.

Direct transfer of a zero mode anyon $c_{\text{zm}}$ between the endpoint defects is physically distinct from zero mode tunneling labeled by $a$. The two are related through $c_{\text{zm}}\in a\times a$, but this relation does not make the corresponding tunneling processes equivalent. This process is the defect analogue of direct topological-charge tunneling between two anyons in Ref.~\cite{bonderson-2009-split}. Its fusion-channel dependence is determined by a $G$-crossed $F$-symbol, a matrix element relating fusion bases with different groupings of the anyons and defects \cite{barkeshli-2019} (see details in App.~\ref{subapp:G-F}). Denoting the direct-transfer operator by $T_{c_{\rm zm}}$, its action in the defect fusion-channel basis is
\begin{align}
    T_{c_{\text{zm}}}\ket{c}=\sbr{F_{c}^{X_{\vb{C}}c_{\text{zm}}\bar{X}_{\vb{C}}}}_{X_{\vb{C}}\bar{X}_{\vb{C}}}\ket{c}\equiv\F(c)\ket{c}\,.\label{eqn:zm_f}
\end{align}
Here the external labels of the $F$-symbol are $X_{\vb{C}}$, $c_{\text{zm}}$ and $\bar{X}_{\vb{C}}$; the lower label $c$ is their total fusion outcome; and the two intermediate channels are fixed to $X_{\vb{C}}$ and $\bar{X}_{\vb{C}}$ by the zero mode conditions
\begin{align}
    X_{\vb{C}}\times c_{\text{zm}}\ni X_{\vb{C}}\,,\quad \bar{X}_{\vb{C}}\times \overline{c_{\text{zm}}}\ni \bar{X}_{\vb{C}}\,.
\end{align}
The $F$-symbol in Eq.~\eqref{eqn:zm_f} depends on the choice of fusion/splitting space basis, whereas the physical tunneling contribution is basis independent. Under a change of basis, the coefficient representing the microscopic tunneling process acquires a compensating factor, leaving its product with the $F$-symbol unchanged. Consequently, the observable content lies in the fusion-channel dependence of the resulting energy shifts, rather than in the isolated $F$-symbol.

In the following sections, we apply this procedure to FQH-SCs with several parent FQH states of interest and derive the spectra associated with the two tunneling processes. For the Abelian and Moore–Read FQH-SCs, the two processes have the same fusion-channel dependence up to a channel-independent normalization. For the Read–Rezayi FQH-SC, however, they produce different relative tunneling amplitudes between the neutral fusion channels \footnote{In Ref.~\cite{bonderson-2009-split}, the monodromy process is more suppressed than direct topological-charge transfer at large separations because the virtual anyon traverses a longer path. Here, the monodromy loop represents zero mode tunneling between FQH-SC segments, as described by Eq.~\eqref{eqn:tunnel}. We treat zero mode tunneling and direct zero mode anyon transfer as separate effective couplings within the defect fusion space, retaining each at leading order in its respective tunneling amplitude. The relative magnitudes of these amplitudes depend on the microscopic junction.}.

\section{Abelian FQH-SC: \texorpdfstring{$\nu=1/m$}{1/m} Laughlin FQH-SC}\label{sec:1/m}
The $\nu=1/m$ Laughlin FQH-SC provides an Abelian benchmark for the construction developed in Sec.~\ref{sec:setup}. The Abelian fusion rules allow the zero mode anyon criterion, bulk-to-boundary projection, and tunneling response to be evaluated explicitly. The construction recovers the parafermion zero mode algebra \cite{fendley-2012}, distinguishes the zero mode structures of fermionic and bosonic Laughlin states \cite{barkeshli-2013}, and reproduces the fractional Josephson effect \cite{clarke-2013,lindner-2012,cheng-2012}. Additional Abelian examples are discussed in App.~\ref{app:abelian}.

The Laughlin FQH state at filling fraction $\nu=1/m$ has $m$ topologically distinct quasiparticle sectors modulo fusing with local physical particles, labeled by $j\in\Z_m$. In the notation of Ref.~\cite{bonderson-2007-thesis}, its anyon theory is $U(1)_m\simeq \Z_m^{(1/2)}$. The sector $j$ carries electric charge $q_j=je/m$, and the modular $\S$-matrix of the anyon theory $\Z_N^{(p)}$ is 
\begin{align}
     \S^{\Z_N^{(p)}}_{jk}=\frac{1}{\sqrt{N}}e^{i \frac{4p\pi}{N}jk}\,.
\end{align}
The parameter $p$ determines the mutual braiding and will be important when comparing the Laughlin state with the charge sector of the Read–Rezayi state in Sec.~\ref{subsec:RR}. For a fermionic Laughlin FQH state, we evaluate the Wilson algebra in the minimal modular extension $U(1)_{4m}$ \footnote{This modular extension can be understood by viewing the fermionic $U(1)_m$ theory as the $\Z_2$ quotient $U(1)_m\simeq U(1)_{4m}/\Z_2$. The quotient treats the emergent fermion of $U(1)_{4m}$ as the local physical electron and removes sectors that braid nontrivially with it. Reinstating these sectors yields the minimal modular extension $U(1)_{4m}$.} to incorporate the fermion-parity-flux sectors. In this extension, the anyons labeled by $r\in\Z_{4m}$ carry electric charge $q_r=re/2m$, and the physical electron $\Psi$ is labeled by $r=2m$. Even $r$ describe the physical quasiparticle sectors, including the electron, whereas odd $r$ describe fermion-parity-flux sectors.

Within the FQH-SC segment, density-density interactions between the counter-propagating Laughlin edges renormalize the mode velocity $u$ and Luttinger parameter $K$, while Cooper pairing generates a cosine perturbation that can gap the edges. The resulting FQH-SC theory takes the form of the sine-Gordon model \cite{clarke-2013,cheng-2012,lindner-2012,schiller-2023,cao-2024} (see App.~\ref{subapp:Lrg} for a detailed derivation):
\begin{align}
    L_{1/m,\De}=\int_{0}^L\rd x\,\cbr{\frac{4m}{2\pi}\frac{1}{K}\sbr{\frac{1}{u}\pbr{\pa_t\vt}^2-u\pbr{\pa_x\vt}^2}+\De\cos(4m\vt)}\,\label{eqn:1/m}.
\end{align}
Here $\vp,\vt$ are nonchiral bosonic fields defined by $\vp,\vt=(\phi_+\pm\phi_-)/2$ with $\phi_\pm$ being the bosonic fields on the two counter-propagating Laughlin edges. They satisfy the equal-time commutation relation
\begin{align}
    \comm{\vp(x_1)}{\vt(x_2)}=i\frac{\pi}{4m}\Th(x_1-x_2)\,,
\end{align}
where $\Th(x)$ is the Heaviside step function. The pairing perturbation has scaling dimension $mK$ and is relevant at weak coupling for $K<2/m$. We consider the resulting gapped phase, in which $\vt$ is pinned near a minimum of the cosine potential.

For the fermionic Laughlin FQH-SC, the bulk-to-boundary projection maps the Wilson operators labeled in $U(1)_{4m}$ to the zero mode operator $\al_1$ and projected loop operator $U_1$, respectively. At the right endpoint, we place the projected loop operator at $x=L$ and define the zero mode operator within a localized region of length $\xi_1$ inside the segment \cite{cao-2024}:
\begin{align}
    \al_1=\frac{1}{\xi_{1}}\int_{L-\xi_1}^{L}\rd x\,e^{i2\vt(x)}\approx e^{i2\vt(L)}\,,\quad U_1(x=L)=e^{i2\vp(L)}\,.\label{eqn:LaughlinU}
\end{align}
The approximate local expression follows in the strong-coupling limit, when fluctuations of $\vt$ are negligible over the localization length $\xi_1$. The fusion rule $1\times 1=2$ identifies the zero mode anyon associated with $\al_1$ as $c_{\text{zm}}=2$. $\al_{1}$ and $U_{1}$ satisfy the $\Z_{2m}$ parafermion zero mode algebra \cite{fendley-2012}. 

In addition, $U_{1}$ generates the $\Z_{2m}$ shift symmetry of the FQH-SC theory \footnote{This is a $\Z_{2m}$ symmetry because the physical objects are the quasiparticle operators $\psi_{qp,\pm}\sim e^{\pm i 2\phi_\pm}$ and the $U_{1}$ symmetry action is $2m$-fold on the quasiparticle operators.}:
\begin{align}
    U_1^\da \phi_\pm U_1=\phi_\pm\pm\frac{\pi}{2m}\,,\quad U_1:\vt\ra\vt+\frac{\pi}{2m}\,.\label{eqn:z2m}
\end{align}
It also anticommutes with the chiral and anti-chiral electron operators,
\begin{align}
    U_1^\da \Psi_{\pm} U_1=-\Psi_\pm\,.
\end{align}
On local electron operators, \(U_1\) therefore acts as the total fermion-parity operator $(-1)^F\equiv(-1)^{F_+}(-1)^{F_-}$ \footnote{Here $(-1)^{F_\pm}$ only anticommutes with the chiral/anti-chiral electron operator $\Psi_\pm$ while commuting with the other electron operator.}. The zero mode operator $\al_1$ transfers electric charge $e/m$, consistent with its transformation under the $\Z_{2m}$ symmetry generated by $U_1$. Although the sine-Gordon model in Eq.~\eqref{eqn:1/m} may possess a larger shift symmetry, only the $\Z_{2m}$ subgroup generated by $U_1$ descends from a bulk Wilson operator. The additional formal transformations therefore do not enlarge the topological ground-state degeneracy or the dimension of the defect Hilbert space.

The tunneling of zero mode operators determines the Josephson spectrum of a junction formed by two fermionic Laughlin FQH-SCs. The leading process is represented by the zero mode tunneling operator $\al_1^{(1)\da}\al_1^{(2)}$. The corresponding zero mode anyon is labeled by $c_{\text{zm}}=2$. We label the fusion channel of the two defects by $c=2l$, where $l\in\Z_{2m}$. Using the modular $\S$-matrix of $U(1)_{4m}$, the monodromy of the bulk Wilson line labeled by $a=1$ around the total topological charge $c=2l$ is
\begin{align}
    M_{ac}=M_{1,2l}^{U(1)_{4m}}=e^{\frac{2\pi i}{4m}2l}\,.
\end{align}
Direct transfer of the zero mode anyon $c_{\text{zm}}=2$ is governed by the $G$-crossed $F$-symbol introduced in Sec.~\ref{subsec:tunnel}. In the gauge used in Ref.~\cite{tambara-1998}, the relevant matrix element is
\begin{align}
    \sbr{F_{2l}^{X_{\vb{C}}2\bar{X}_{\vb{C}}}}_{X_{\vb{C}}\bar{X}_{\vb{C}}}=e^{\frac{2\pi i}{4m}2l}\,.
\end{align}
Thus, zero mode tunneling and direct zero mode anyon transfer have the same fusion-channel dependence in the Abelian Laughlin case, up to a channel-independent normalization. Absorbing this normalization into $\Ga_1$ and shifting the superconducting phase difference $\phi_{\rm SC}$ to absorb the channel-independent tunneling phase, we obtain the zero mode tunneling spectrum
\begin{align}
    E_l(\phi_{\text{SC}})=2\abs{\Ga_1}\cos(\frac{2\pi l}{2m}-\frac{\phi_{SC}}{2m})\,.
\end{align}
Each fixed fusion-channel branch is $4m\pi$-periodic in $\phi_{\text{SC}}$, reproducing the fractional Josephson effect of a $\Z_{2m}$ parafermion zero mode \cite{kitaev-2001,clarke-2013,cheng-2012,lindner-2012}. 

For a bosonic Laughlin state at filling fraction $\nu=1/m$ with $m$ even, the zero mode anyons form the even subgroup $2\Z_m\simeq\Z_{m/2}$, as follows from the defect fusion rules in App.~\ref{subapp:bZm}. This gives the intrinsic $\Z_{m/2}$ zero mode structure \cite{barkeshli-2013}. This difference illustrates that the local edge expressions alone do not determine the complete zero mode structure; the fermionic modular extension and the resulting Wilson algebra must also be taken into account.

\section{FQH-SCs with non-Abelian parent states}\label{sec:nonab}
Although it is well established that FQH-SCs built from Abelian states can support non-Abelian defects \cite{barkeshli-2013,barkeshli-2013-2}, their zero mode algebra retains a group-like, parafermionic structure \cite{fradkin-1980}. For a non-Abelian parent state, the bulk-to-boundary construction incorporates non-Abelian fusion rules and Wilson-line actions of the parent topological order, which may be non-invertible on the defect Hilbert space \cite{chang-2019}. In particular, the fusion product entering the zero mode anyon criterion in Eq.~\eqref{eqn:zm_def} can contain several outcomes:
\begin{align}
    a\times a=\sum_{c_{\text{zm}}}N_{aa}^{c_{\text{zm}}}\,c_{\text{zm}}\,.
\end{align}
A single Wilson label $a$ can therefore correspond to several candidate zero mode anyons. The supported outcomes are selected by the defect-absorption condition in Eq.~\eqref{eqn:Gzm_def}, and may include both charged and electrically neutral anyons.
\subsection{Generalized Moore-Read FQH-SC} \label{sec:MR}
The generalized Moore–Read state at the filling fraction $\nu=1/m$ \cite{moore-1991,fendley-2007} provides the simplest realization of the non-Abelian fusion structure described above. The $m=2$ case was proposed to be a candidate topological order for the fractional component of the $\nu=5/2$ FQH state \cite{moore-1991}. The MR state contains a neutral Ising sector and a charged $U(1)_{4m}$ sector. The Ising sector contains the vacuum $1$, the Majorana fermion $\chi$ and the non-Abelian Ising anyon $\si$, which has quantum dimension $d_\si=\sqrt{2}$. Their fusion rules are
\begin{align}
    \chi\times\chi=1\,,\quad \chi\times\si=\si\,,\quad \si\times\si=1+\chi\,.\label{eqn:Ising}
\end{align}
The modular $\S$-matrix in the $\cbr{1,\si,\chi}$ basis is
\begin{align}
    \S^{\text{Ising}}=\frac{1}{2}\mqty(1&\sqrt{2}&1\\\sqrt{2}&0&-\sqrt{2}\\1&-\sqrt{2}&1)\,.
\end{align}
The anyons in the MR state are labeled $1_l$, $\si_l$, and $\chi_l$, with $l\in\Z_{4m}$ and the physical electron (or physical boson for bosonic MR states) labeled by $\Psi\equiv\chi_{2m}$. The locality with respect to the electron requires $l$ to be even for $1_l$ and $\chi_l$, and odd for $\si_l$. We focus on the fermionic case first and later briefly discuss the bosonic case with odd $m$. For fermionic MR FQH-SC, a convenient choice of minimal modular extension is $\text{Ising}\boxtimes U(1)_{4m}$ which in addition includes sectors that have monodromy -1 with the electron $\Psi$. We label sectors in the minimal modular extension by $1_r$, $\si_r$, and $\chi_r$, with no constraint on $r\in\Z_{4m}$.

Applying the zero mode anyon criterion in Eq.~\eqref{eqn:zm_def} to the Wilson labels $a=1_r,\chi_r,\si_r$ gives
\begin{align}
    1_r\times 1_r=1_{2r}\,,\quad \chi_r\times \chi_r=1_{2r}\,,\quad \si_r\times \si_r=1_{2r}+\chi_{2r}\,.
\end{align}
The fusion outcomes yield a charged PZM sector generated by the anyon $1_2$, which carries electric charge $e/m$, together with the neutral candidate $\chi_0$ from $\si_0\times\si_0=1_0+\chi_0$. The supported candidates depend on the defect type through the condition $X_{\vb{C}}\times c_{\text{zm}}\ni X_{\vb{C}}$. As shown in App.~\ref{subapp:fMRdefect}, the defects $X_{\si_0}$ and $X_{\si_1}$ support both the charged and neutral zero mode anyons. We refer to these as $X_\si$-type defects and focus on this choice of defect type. Their zero mode structure is $\Z_2^{(n)}\times\Z_{2m}$, where $\Z_2^{(n)}=\cbr{1_0,\chi_0}$ denotes the neutral zero mode anyon sector.

The local realization of the charged and neutral candidates follows from the $(1+1)$-dimensional edge theory of an MR FQH-SC. The two counter-propagating edges contain right- and left-moving Majorana fields $\chi_\pm$, describing the neutral Ising sector, and bosonic fields $\phi_\pm$, describing the $U(1)_{4m}$ charge sector \cite{milovanovic-1996}: 
\begin{align}
    \L_{\text{MR},\pm}=\frac{i}{2}\chi_\pm (\pa_t\pm v_n\pa_x)\chi_\pm+\frac{4m}{4\pi}\pa_x\phi_\pm(\mp\pa_t-v_c\pa_x)\phi_\pm\,,\label{eqn:MR_bdy}
\end{align}
where $v_n,v_c>0$ are velocities of the Majorana fermion and the charged boson, respectively. These fields are Hermitian, $\chi_\pm^\da=\chi_\pm,~\phi^\da_\pm=\phi_\pm$ and obey the canonical equal-time relations
\begin{gather}
    \comm{\phi_\pm(x_1)}{\pa_{x_2}\phi_\pm(x_2)}=\mp i\frac{2\pi}{4m}\de(x_1-x_2)\,,\quad\comm{\phi_+(x_1)}{\phi_-(x_2)}=-i\frac{\pi}{4m}\,,\nonumber\\
    \acomm{\chi_\pm(x_1)}{\chi_\pm(x_2)}=\de(x_1-x_2)\,,\quad \acomm{\chi_+(x_1)}{\chi_-(x_2)}=0\,.
\end{gather}
Within the FQH-SC segment, density-density interactions are encoded in the renormalized charge-mode velocity $u$ and Luttinger parameter $K$, while a relevant proximity-induced Cooper pairing gaps both the neutral and charge sectors \cite{sohal-2020}. The MR FQH-SC theory can be written in terms of a two-component Majorana spinor $\chi=(\chi_+,\chi_-)^T$ and the nonchiral boson $\vt=(\phi_+-\phi_-)/2$ as
\begin{align}
    \L_{\text{MR},\De}=\frac{i}{2}\bar{\chi}\sl{\pa}\chi+\frac{4m}{2\pi}\frac{1}{K}\sbr{\frac{1}{u}\pbr{\pa_t\vt}^2-u\pbr{\pa_x\vt}^2}-\frac{\De}{2}\bar{\chi}\chi\cos(4m\vt).\label{eqn:MRD}
\end{align}
We have set the neutral-mode velocity to unity and define $\bar{\chi}\equiv\chi^\da\ga^0$, $\sl{\pa}\equiv\ga^\mu\pa_\mu$ with $\ga^0=\si^y$ and $\ga^1=+i\si^x$ where $\si^i$ are the Pauli matrices. The pairing operator has scaling dimension $1+mK$ and is relevant at weak coupling for $K<1/m$; its RG flow is discussed in App.~\ref{subapp:MRrg}. In the resulting gapped phase, localized zero mode operators may involve charge-sector vertex operators, neutral Ising primary fields, or products of the two, as determined by their bulk Wilson labels.

The bulk-to-boundary projection connects the Wilson labels $1_1$ and $\si_0$ to local operators that access the charged and neutral zero mode sectors. On the two counter-propagating edges, the anyon label $1_1$ is represented by charge-sector vertex operators $e^{\pm i\phi_\pm}$, while the anyon label $\si_0$ is represented by the Ising spin fields $\si_\pm$. We use the right-endpoint convention of Sec.~\ref{sec:1/m}, placing the projected loop operators at $x=L$ and localizing the zero mode operators immediately inside the segment.
\begin{align}
    \al_{1_1}&=\frac{1}{\xi_1}\int^{L}_{L-\xi_1}\rd x~e^{i2\vt(x)}\approx e^{i2\vt(L)}\,,\\
    \al_{\si_0}&=\frac{1}{\xi_\si}\int^L_{L-\xi_{\si}}\rd x~\si_+(x)\si_-(x)\approx \si_+(L)\si_-(L)\,.
\end{align}
Here $\xi_1$ and $\xi_\si$ are the corresponding localization lengths. The fusion rule $1_1\times1_1=1_2$ identifies $1_2$ as the zero mode anyon corresponding to $\al_{1_1}$, while $\si_0\times\si_0=1_0+\chi_0$ identifies $\chi_0$ as the neutral zero mode anyon corresponding to $\al_{\si_0}$. The fusion rule $\chi_0\times\chi_0=1_0$ supplies no additional nontrivial zero mode anyon. A zero mode operator sensitive to both sectors is $\al_{\si_1}$, which combines the neutral Ising spin field with the elementary charge-sector vertex operator.

The projected loop operators encode both the charge and neutral actions on the localized zero modes:
\begin{align}
    U_{1_1}=e^{i2\vp(L)}\,,\quad U_{\chi_0}\sim i\chi_+(L)\chi_-(L)\,,\quad U_{\si_0}\sim\si_+(L)\si_-(L)\,,
\end{align}
where $\varphi\equiv(\phi_++\phi_-)/2$. The charge sector operator $U_{1_1}$ generates the $\Z_{2m}$ charge-sector symmetry and acts on the charged zero mode operator as
\begin{align}
    U_{1_1}^\da \al_{1_1}U_{1_1}=e^{\frac{2\pi i}{2m}}\al_{1_1}\,.
\end{align}
The remaining two operators encode the invertible and non-invertible Wilson-line actions of the neutral Ising sector. The operator $U_{\chi_0}$ implements the invertible $\chi_0$ Wilson-line action and leaves $\al_{\si_0}$ invariant. The $\si_0$ Wilson line instead gives the non-invertible action
\begin{align}
    \U_{\si_0}[\al_a]: \al_{1_1}\mapsto2\al_{1_1}\,,\quad\al_{\si_0}\mapsto0\,.
\end{align}
These projected loop actions act within the neutral zero mode sector already determined by defect fusion rules. Despite sharing the same local primary-field content, $U_{\si_0}$ and $\al_{\si_0}$ descend from distinct Wilson configurations and are distinguished by their spatial ordering: $U_{\si_0}$ is placed at the endpoint, whereas $\al_{\si_0}$ is localized slightly inside the FQH-SC segment. 

Both tunneling processes introduced in Sec.~\ref{subsec:tunnel} probe the neutral $\Z_2^{(n)}$ sector. For fermionic MR states, $c=(l_n,2l)$ denotes the physical fusion channel $1_{2l}$ for $l_n=0$ and $\chi_{2l}$ for $l_n=1$, with $l\in\Z_{2m}$. Because the modular data of the chosen minimal modular extension factorize into Ising and $U(1)_{4m}$ sectors, the monodromy factor also factorizes. The zero mode operator tunneling then probes the monodromy between $a=\si_1$ and $c=(l_n,2l)$:
\begin{align}
    M_{ac}=M_{\si_1,(l_n,2l)}=M_{\si,l_n}^{\text{Ising}}M_{1,2l}^{U(1)_{4m}}=(-1)^{l_n}e^{\frac{2\pi i}{4m}2l}\,,
\end{align}
where $M_{\si,1}^{\text{Ising}}=1$ and $M_{\si,\chi}^{\text{Ising}}=-1$ are the relevant Ising monodromy factors. Thus, the neutral fusion channel contributes only a relative sign, whereas the charge fusion channel exhibits the same fractional phase winding as in the Abelian PZM case.

For the factorized defect theory used here, the relevant $G$-crossed $F$-symbol separates into a neutral Ising $F$-symbol and a charge-conjugation defect matrix element. For comparison with tunneling of $\al_{\si_1}$, we consider direct transfer of $c_{\rm zm}=\chi_2$, the outcome carrying both the neutral and charge labels.
\begin{align}
    \sbr{F_{(l_n,2l)}^{X_{\si}c_{\text{zm}}\bar{X}_{\si}}}_{X_{\si}\bar{X}_{\si}}=\sbr{F_{l_n}^{\si\chi\si}}_{\si\si}\sbr{F_{2l}^{X_{0}2\bar{X}_{0}}}_{X_{0}\bar{X}_{0}}=(-1)^{l_n}e^{\frac{2\pi i}{4m}2l}\,,
\end{align}
where $\sbr{F_{1}^{\si\chi\si}}_{\si\si}=1$ and $\sbr{F_{\chi}^{\si\chi\si}}_{\si\si}=-1$ are the relevant $F$-symbols in the Ising sector \cite{kitaev-2006,bonderson-2007-thesis} and $X_0$ is the charge-conjugation defect in the charge sector. Thus, tunneling of the zero mode operator and direct transfer of the zero mode anyon have the same fusion-channel dependence, up to a channel-independent normalization. As in the Laughlin case, this normalization can be absorbed into the corresponding nonuniversal tunneling amplitude. 

Since tunneling of $\al_{\si_1}$ transfers charge $e/m$, the tunneling amplitude acquires a superconducting phase $e^{-i\phi_{\text{SC}}/(2m)}$ from gauge invariance. Absorbing the channel-independent normalization into the tunneling amplitude $\Ga_\si$, the Josephson spectrum is
\begin{align}
    E_{(l_n,2l)}(\phi_\text{SC})=2\abs{\Ga_\si}\cos(\frac{2\pi}{2}l_n+\frac{2\pi}{2m}l-\frac{\phi_{\text{SC}}}{2m})\,.
\end{align}
Each fixed fusion-channel branch is $4m\pi$-periodic in $\phi_\text{SC}$. The two neutral fusion channels satisfy $E_{(1,2l)}(\phi_\text{SC})=-E_{(0,2l)}(\phi_\text{SC})$ and therefore differ by a relative $\pi$ phase shift while having equal tunneling magnitudes.

For bosonic MR FQH-SC at filling fraction $\nu=1/m$ with odd $m$, the identification of anyons modulo the physical boson changes the allowed zero mode operators. The relevant zero mode operators here are $\al_{1_2}$ and $\al_{\si_1}$, while the projected ground-state operators are $U_{1_2}$, $U_{\chi_0}$ and $U_{\si_1}$. The resulting zero mode structure is $\Z_2^{(n)}\times\Z_m$. 

The construction above provides localized operator realizations of the charged and neutral zero modes within the $\mathbb Z_2^{(n)}\times\mathbb Z_N$ structure. For fermionic MR FQH-SCs, the two neutral fusion channels contribute opposite signs but equal magnitudes to the tunneling spectrum, and the two tunneling processes considered here have the same fusion-channel dependence. These results underlie the Josephson and interferometric signatures studied in Ref. \cite{cao-2026-l}. In a setup consisting of an FQH-SC Josephson junction with a single QPC, tunneling of the Ising anyon $\si_1$ suppresses the tunneling of zero mode operator $\al_{\si_1}$. In addition, quasiparticle tunneling produces stochastic transitions between spectral branches associated with tunneling of $\al_{1_1}$: tunneling of $\si_1$ shifts the charge-sector label by one, whereas tunneling of $1_2$ shifts it by two. In the two-QPC Fabry–P\'{e}rot configuration, weak zero mode tunneling allows the defect fusion channel to be probed through interference. Strong zero mode tunneling instead selects a different fusion basis, producing a superposition of the channels enclosed by the interferometer and exact cancellation of the interference term in the topological limit \cite{cao-2026-l}.


\subsection{States at \texorpdfstring{$\nu=1/2$}{1/2}}
The Moore–Read result above provides a basis for comparing the principal candidate topological orders at filling fraction $\nu=1/2$, including the Moore–Read Pfaffian, anti-Pfaffian (A-Pf) \cite{levin-2007,lee-2007}, particle-hole-symmetric Pfaffian (PHPf) \cite{son-2015}, and Abelian Halperin $(331)$ state \cite{halperin-1983,read-1996}. Comparing their FQH-SC realizations reveals which of these distinctions survive the construction.

The MR, A-Pf, and PHPf states all contain an Ising-type neutral sector with fusion rules in Eq.~\eqref{eqn:Ising}. Their distinctions arise primarily from the relative chiralities and topological spins of the neutral and charge sectors. For the corresponding endpoint defects, the neutral monodromy factors entering the tunneling calculation are $M_{\si,1}=1$ and $M_{\si,\chi}=-1$ in all three states, while the charge sector gives the same phase winding. Consequently, they all support the same $\Z_2^{(n)}\times\Z_4$ zero mode structure and the same fusion-channel dependence for the tunneling processes considered here: $8\pi$-periodic branches and a relative $\pi$ shift between neutral channels of equal magnitude. These zero mode properties cannot distinguish the MR, A-Pf, and PHPf states. Doing so requires probes sensitive to edge chirality, chiral central charge, or topological spin.

The Halperin $(331)$ state provides a contrasting Abelian candidate. It is described by
\begin{align}
    K=\mqty(3&1\\1&3)\,,\quad t=\mqty(1\\1)\,.
\end{align}
Our $K$-matrix conventions are reviewed in App.~\ref{app:kmat}. The Abelian anyon group of the Halperin $(331)$ state is $\A_{331}=\Z^2/K\Z^2\simeq\Z_8$ \cite{wen-1992}. As shown in App.~\ref{subsubsec:fAb}, the FQH-SC supports a $\Z_4$ zero mode structure and an $8\pi$-periodic fractional Josephson effect. Unlike the Pfaffian-type states, the Halperin $(331)$ FQH-SC does not contain an independent neutral Ising fusion sector.

\subsection{Read-Rezayi FQH-SC at \texorpdfstring{$\nu=3/(3M+2)$}{3/(3M+2)}}\label{subsec:RR}
The $k=3$ Read–Rezayi states \cite{read-1999} provide a non-Abelian example with a Fibonacci neutral sector that supports universal TQC \cite{freedman-2000}. These states occur at the filling fraction $\nu=3/(3M+2)$ \footnote{For generic $k$, the parent state corresponds to the $(U(2)_{k,-2k}\times U(1)_{k(kM+2)})/\Z_k$ CS theory with filling fraction $\nu=k/(kM+2)$, where $U(2)_{k,-2k}\simeq \frac{SU(2)_k}{U(1)_{2k}}$ is the $\Z_k$ parafermion theory defined via Goddard–Kent–Olive coset construction \cite{goddard-1985}. For $k=2$, this gives the Moore-Read states discussed in the previous subsection with identification $m=M+1$.}. The $k=3$ RR states are fermionic when $M$ is odd and bosonic when $M$ is even. In particular, the state with $\nu=3/5$ has been proposed as a candidate topological order for the fractional component of the $\nu=13/5$ FQH state, while its particle-hole conjugate has been proposed for the $\nu=12/5$ state \cite{read-1999,sreejith-2013,zhu-2015,pakrouski-2016,mong-2017}. Its edge theory contains a neutral sector described by the $\Z_3$ parafermion CFT \cite{read-1999,zamolodchikov-1985} and a charged chiral-boson sector. 

Combining these sectors, imposing locality with respect to the local physical particle, and identifying excitations that differ by fusion with the local particle yields $2(3M+2)$ physical anyon sectors. A convenient way to organize these anyons is decomposing each anyon into a charged and a neutral sector. The resulting anyon theory is $\C_{\text{fRR}}\simeq \text{Fib}\boxtimes \Z_{3M+2}^{\pbr{(M+1)/2}}$ \cite{bonderson-2007-thesis,cheng-2026-order} for fermionic RR states and $\C_{\text{bRR}}\simeq \text{Fib}\boxtimes \Z_{3M+2}^{\pbr{2M+3/2}}$ for bosonic RR states (see derivation in App. \ref{subapp:bRRdefect}), where $\text{Fib}\equiv\{1,\tau\}$ is the Fibonacci topological order. We label the anyons by $1_j$ and $\tau_j$, where $\tau$ denotes the Fibonacci anyon and $j\in\Z_{3M+2}$ labels the charge sector. Their fusion rules are
\begin{align}
    1_j\times1_{j'}=1_{j+j'}\,,\quad 1_j\times \tau_{j'}=\tau_{j+j'}\,,\quad \tau_j\times\tau_{j'}=1_{j+j'}+\tau_{j+j'}\,.
\end{align}
The Fibonacci anyon has quantum dimension equal to the golden ratio $\Phi$
\begin{align}
    d_\tau=\Phi\equiv\frac{1+\sqrt{5}}{2}\,,
\end{align}
and the modular $\S$-matrix of the Fibonacci sector, in the $\cbr{1,\tau}$ basis, is
\begin{align}
    \S^{\text{Fib}}=\frac{1}{\sqrt{1+\Phi^2}}\mqty(1&\Phi\\\Phi&-1)\,.
\end{align}

For fermionic RR FQH-SC, we employ a minimal modular extension when applying the zero mode criterion and bulk-to-boundary projection. A convenient choice is $\text{Fib}\boxtimes \Z_{4(3M+2)}^{(3/2)}$ (see details in App.~\ref{subapp:RRdefect}). We then label sectors in the modular extension by $1_r$ and $\tau_r$ with $r\in\Z_{4(3M+2)}$. Equation \eqref{eqn:zm_def} gives
\begin{align}
    1_{r}\times 1_{r}=1_{2r}\,,\quad \tau_r\times\tau_r=1_{2r}+\tau_{2r}\,.
\end{align}
The anyon $1_2$ generates the $\Z_{2(3M+2)}$ PZM charge sector, whereas $\tau_0$ is electrically neutral. As discussed in App.~\ref{subapp:RRdefect}, the defect $X_\tau=X_0\times\tau_0$, obtained by attaching the neutral Fibonacci anyon to the bare charge-conjugation defect $X_0$, supports both the charged and neutral zero mode anyons. We focus on this defect type, whose zero mode structure is $\mathrm{Fib}\times\mathbb Z_{2(3M+2)}$.

The localized operators corresponding to these zero mode anyons follow from the $(1+1)$-dimensional edge theory of the RR FQH-SC. Each edge contains a neutral $\Z_3$ parafermion sector and a $U(1)_{12(3M+2)}$ charge sector \cite{zamolodchikov-1985,read-1999,cappelli-2001},
\begin{align}
    \L_{\text{RR},\pm}=\L_{\Z_3,\pm}+\frac{12(3M+2)}{4\pi}\pa_x\phi_\pm(\mp\pa_t-v_c\pa_x)\phi_\pm\,.
\end{align}
The fermion-parity-flux sectors introduced by the minimal modular extension correspond to primary fields with nontrivial monodromy with the electron operator (or charge-$e$ simple current)
\begin{align}
    \Psi_\pm=\psi_{1,\pm}e^{\pm 2i (3M+2)\phi_\pm}\,,
\end{align}
where $\psi_{1,\pm}$ are the chiral and anti-chiral  parafermion fields of the $\Z_3$ CFT. The CFT describing this extension is obtained by extending the product chiral algebra by the bosonic charge-$2e$ simple current 
\begin{align}
    J_{2e,\pm}=\psi_{2,\pm}e^{\pm 4i(3M+2)\phi_\pm}\,.
\end{align}
Its primary sectors are obtained by retaining fields with trivial monodromy with $J_{2e,\pm}$ and identifying those related by fusion with this current.

Including density-density interactions and the Cooper-pairing perturbation $\De\Psi_+\Psi_-+\hc$ \footnote{Klein factors ensuring the fermionic statistics of the electron operators are implicit, and their phase has been absorbed into $\De$.}, the RR FQH-SC theory may be written in terms of the nonchiral charge field $\vt\equiv(\phi_+-\phi_-)/2$ as
\begin{align}
    \L_{\text{RR},\De}=\L_{\Z_3,+}+\L_{\Z_3,-}+\frac{12(3M+2)}{2\pi}\frac{1}{K}\sbr{\frac{1}{u}(\pa_t\vt)^2-u(\pa_x\vt)^2}+\De\psi_{1,+}\psi_{1,-}e^{4(3M+2)i\vt}+\hc\,,\label{eqn:RRD}
\end{align}
where $u$ and $K$ are the renormalized charge velocity and Luttinger parameter. The pairing operator has scaling dimension $4/3+(3M+2)K/3$ and is relevant at weak coupling for $K<2/(3M+2)$ (see details in App.~\ref{subapp:RRrg}).

The projection of Sec.~\ref{sec:setup} expresses the localized zero mode operators in terms of $\Z_3$ parafermion CFT primary fields \cite{zamolodchikov-1985,cappelli-2001} and charge-sector vertex operators. The fermion parity flux $1_1$ corresponds to the primary field $e^{\pm 3i\phi_\pm}$. The neutral Fibonacci anyon $\tau_0$ is represented by the primary field $\ep_\pm$. The corresponding endpoint zero mode operators are therefore
\begin{align}
    \al_{1_1}&=\frac{1}{\xi_1}\int_{L-\xi_1}^{L}\rd x~e^{6i\vt(x)}\approx e^{6i\vt(L)}\,,\\
    \al_{\tau_0}&=\frac{1}{\xi_\ep}\int_{L-\xi_\ep}^{L}\rd x~\ep_{+}(x)\ep_{-}(x)\approx\ep_{+}(L)\ep_{-}(L)\,,
\end{align}
where $\xi_1$ and $\xi_\ep$ are the corresponding localization lengths. The zero mode operator $\al_{1_1}$ carries charge $3e/(3M+2)$ and the fusion outcome $1_1\times 1_1=1_2$ identifies $1_2$ as the charged zero mode anyon corresponding to $\al_{1_1}$. Similarly, $\tau_0\times\tau_0=1_0+\tau_0$ identifies $\tau_0$ as the neutral zero mode anyon corresponding to $\al_{\tau_0}$. The zero mode operator sensitive to both sectors is $\al_{\tau_1}$, whose Wilson label gives $\tau_1\times\tau_1=1_2+\tau_2$. The outcome $\tau_2$ carries both the neutral and charged labels.

The corresponding projected loop operators are
\begin{align}
    U_{1_1}=e^{6i\vp(L)}\,,\quad U_{\tau_0}=\ep_{+}(L)\ep_{-}(L)\,,
\end{align}
where $\vp\equiv(\phi_++\phi_-)/2$. Although $\al_{\tau_0}$ and $U_{\tau_0}$ involve the same local primary-field product, they descend from different processes and are distinguished by their spatial support and ordering relative to the defect line. Their actions follow from Eq.~\eqref{eqn:zmalg}:
\begin{align}
    \U_{1_1}[\cdot]:~\al_{1_1}\ra e^{\frac{3\pi i}{3M+2}}\al_{1_1}\,,&\quad \al_{\tau_0}\ra \al_{\tau_0}\,,\label{eqn:z5alg}
\end{align}
while the non-invertible operator $U_{\tau_0}$ acts as
\begin{align}
    \U_{\tau_0}[\cdot]:~\al_{1_1}\ra (1+\Phi)\al_{1_1}\,,&\quad\al_{\tau_0}\ra (2-\Phi)\al_{\tau_0}\,.\label{eqn:tau_alg}
\end{align}
The projected loop actions distinguish the charged and neutral zero mode operators within the Fib$\times\Z_{2(3M+2)}$ structure fixed by defect fusion. The Fibonacci loop weights these operators by the real factors $1+\Phi$ and $2-\Phi$, respectively.

The $\text{Fib}\times\Z_{2(3M+2)}$ structure produces nontrivial neutral-channel amplitude ratios and distinguishes zero mode tunneling from direct zero mode anyon transfer. We label the fusion channel of two adjacent $X_\tau$-type defects by $c=(l_n,2l)$ where $l_n\in\text{Fib}$ labels the neutral Fibonacci channel and $l\in\Z_{2(3M+2)}$ labels the charge sector. The zero mode tunneling process probes the monodromy between $a=\tau_1$ and $c=(l_n,2l)$. Since the RR modular data factorize into neutral and charge sectors, its monodromy also factorizes:
\begin{align}
    M_{ac}=M^{\text{Fib}}_{\tau,l_n}M^{\Z_{4(3M+2)}^{(3/2)}}_{1,2l}=\mu_{l_n}e^{\frac{3\pi i}{3M+2}l}\,.\label{eqn:RRmonodromy}
\end{align}
From the Fibonacci modular $\S$-matrix, we obtain $\mu_1=1$ and $\mu_\tau=-1/\Phi^2$. After absorbing a channel-independent phase and normalization into $\Ga_{\tau_1}$, the resulting Josephson spectrum is
\begin{align}
    E_{(l_n,2l)}(\phi_{\text{SC}})=2\abs{\Ga_{\tau_1}}\mu_{l_n}\cos(\frac{3\pi}{3M+2}l-\frac{3\phi_{\text{SC}}}{2(3M+2)})\,.
\end{align}
Each fixed fusion-channel branch exhibits a $4(3M+2)\pi/3$-periodic fractional Josephson effect. The spectra in the two neutral fusion channels are related by \cite{bonderson-2009-split}
\begin{align}
    E_{(\tau,2l)}=-\frac{1}{\Phi^2}E_{(1,2l)}\,.
\end{align}
Unlike the Moore–Read case, where the two neutral channels have equal tunneling magnitudes, the Fibonacci channel produces the universal relative magnitude $\Phi^{-2}$.

The RR FQH-SC also distinguishes tunneling of a zero mode operator from direct transfer of a zero mode anyon. For direct transfer of $c_{\text{zm}}=\tau_2$, the fusion-channel dependence is determined by the $G$-crossed $F$-symbol \cite{bonderson-2007-thesis,tambara-1998}
\begin{align}
    \sbr{F_{(l_n,2l)}^{X_\tau \tau_2\bar{X}_\tau}}_{X_\tau\bar{X}_\tau}=\sbr{F_{l_n}^{\tau \tau\tau}}_{\tau\tau}e^{\frac{3\pi i}{3M+2}l}\,,
\end{align}
where $\sbr{F_{1}^{\tau \tau\tau}}_{\tau\tau}=1$ and $\sbr{F_{\tau}^{\tau \tau\tau}}_{\tau\tau}=-1/\Phi$ in a compatible gauge. The direct zero mode anyon transfer amplitudes therefore satisfy
\begin{align}
    \F(\tau,2l)=-\frac{1}{\Phi}\F(1,2l)\,.
\end{align}
The neutral transfer factors are the established Fibonacci $F$-symbol matrix elements \cite{bonderson-2007-thesis}. Here, their application to FQH-SC defects distinguishes direct zero mode anyon transfer from zero mode tunneling, whose monodromy factors give the ratio $\Phi^{-2}$. These ratios characterize the separate tunneling contributions and probe different topological data within the same defect fusion space.

For bosonic RR FQH-SC at filling fraction $\nu=3/(3M+2)$ with even $M$, the identification of anyons modulo the physical boson changes the allowed zero mode operators. The relevant zero mode operators here are $\al_{1_1}$ and $\al_{\tau_0}$, while the projected ground-state operators are $U_{1_1}$ and $U_{\tau_0}$. Here the anyon labels refer to the bosonic parent theory $\C_{\rm bRR}$. The resulting zero mode structure is $\text{Fib}\times\Z_{\frac{3M+2}{2}}$. 

The construction above then provides localized operator realizations of the $\text{Fib}\times\Z_N$ zero modes where $N=2(3M+2)$ for fermionic RR FQH-SC and $N=\frac{3M+2}{2}$ for bosonic RR FQH-SC. Its Fibonacci sector produces different tunneling magnitudes in the two neutral fusion channels and distinguishes zero mode operator tunneling from direct zero mode anyon transfer through the respective ratios in amplitudes. The Josephson and interferometric configurations proposed in Ref.~\cite{cao-2026-l} provide routes to probing this structure. In the MR FQH-SC case, Ising anyon tunneling suppresses the neutral-sector contribution because $\S_{\sigma\sigma}=0$. In contrast, the Fibonacci linking factors in Eq. \eqref{eqn:tau_alg} are nonzero. If quasiparticle tunneling changes the neutral fusion channel, the resulting stochastic transitions can switch between branches with different amplitudes. Similarly, in the weak zero mode tunneling regime of a Fabry–P\'{e}rot interferometer, a tunneling quasiparticle with Fibonacci label $\tau$ probes the enclosed neutral fusion channel through the monodromy factors $M_{\tau,1}=1$ and $M_{\tau,\tau}=-\Phi^{-2}$. At fixed charge sector and quasiparticle tunneling amplitudes, the two channels therefore produce interference contributions to the longitudinal conductance with a relative magnitude $\Phi^{-2}$ and a relative $\pi$ phase shift.

\section{Discussion}\label{sec:dis}
In this work, we generalized the bulk-to-boundary construction of localized zero mode operators from Abelian FQH-SCs to systems with non-Abelian parent topological orders. Starting from the established charge-conjugation defect description of FQH-SC segments, we distinguished zero mode anyons from the zero mode operators that realize them in the one-dimensional FQH-SC theory. By projecting folded bulk Wilson operators onto the FQH-SC endpoints, we expressed the zero mode and projected loop operators in terms of primary fields in the CFTs describing the parent edges. The bulk loop-removal relation determines the projected loop actions on zero mode operators within the defect fusion space. This construction distinguishes tunneling of zero mode operators, governed by the bulk monodromy $M_{ac}$, from direct transfer of a zero mode anyon, governed by the appropriate matrix element of a $G$-crossed $F$-symbol.

For the endpoint defect types considered here, the construction reproduces the established $\Z_{2m}$ Laughlin parafermion structure \cite{clarke-2013,cheng-2012,lindner-2012} and yields, for generalized Moore–Read states at filling fraction $\nu=1/m$, a $\Z_2^{(n)}\times\Z_N$ zero mode structure, where $N=2m$ for fermionic states and $N=m$ for bosonic states. The Moore–Read neutral fusion channels contribute equal tunneling magnitudes with a relative $\pi$ phase shift. The Read–Rezayi FQH-SC at filling fraction $\nu=3/(3M+2)$ supports a $\text{Fib}\times\Z_N$ structure, where $N=2(3M+2)$ for fermionic states and $N=\frac{3M+2}{2}$ for bosonic states. In the RR FQH-SC system, zero mode tunneling and direct zero mode anyon transfer produce the distinct relative magnitudes $\Phi^{-2}$ and $\Phi^{-1}$, respectively, where $\Phi$ is the golden ratio. The tunneling response therefore probes monodromy and associativity data beyond fusion-channel counting.

The construction applies to other fractionalized systems whenever superconducting pairing realizes the corresponding gapped defect line and the endpoint defect type supports the candidate zero mode anyons. In particular, the construction can be adapted to fractional quantum anomalous Hall states \cite{haldane-1988,neupert-2011,tang-2011,sheng-2011,regnault-2011} and to fractional quantum spin Hall systems with helical edges \cite{kane-2005,bernevig-2006,levin-2009}. For a fixed defect theory, additional global symmetries can constrain the allowed gapping and tunneling operators and assign quantum numbers within the defect fusion space. This occurs, for example, in Halperin spin-singlet states and in systems with spin or valley degeneracies, including graphene-based devices \cite{nomura-2006}.

In the Introduction, we noted that braiding the Laughlin and MR FQH-SC defects considered here does not by itself provide universal TQC. In Ref.~\cite{mong-2014}, the authors proposed an array of interacting $\Z_3$ parafermion zero modes in a spin-unpolarized $\nu=2/3$ FQH-SC heterostructure. Starting from chains tuned to parafermion criticality, suitable interchain couplings produce a gapped two-dimensional phase supporting deconfined Fibonacci anyons and universal TQC through braiding. Our RR construction starts from a non-Abelian parent state whose anyon theory already contains a Fibonacci sector. Superconducting proximity coupling introduces endpoint defects while preserving the parent bulk topological order, and our construction provides localized operators acting on their fusion space.

The construction developed here provides the localized zero mode operators and tunneling Hamiltonians underlying the measurement schemes of Ref.~\cite{cao-2026-l}. For the MR case, that work predicts stochastic switching between Josephson branches induced by quasiparticle tunneling through a QPC and suppression of Fabry–P\'{e}rot interference when zero mode tunneling becomes strong. Experimentally, these effects could be investigated by tuning the QPC tunneling and the coupling between adjacent superconducting endpoints while measuring the Josephson response and longitudinal conductance. Comparing the weak and strong zero mode tunneling regimes would test how the coupling selects the defect fusion basis probed by the interferometer.

Our RR construction extends these proposals to defects with a Fibonacci neutral sector. In the weak zero mode tunneling regime, for a tunneling Fibonacci anyon $\tau$, the enclosed neutral channels $1$ and $\tau$ produce interference contributions with a relative magnitude $\Phi^{-2}$ and a relative $\pi$ phase shift, at fixed charge sector and quasiparticle tunneling amplitudes. Measuring both the magnitude and phase of the conductance oscillations would therefore distinguish these channels, whereas the MR neutral channels have equal interference magnitudes. Branch-resolved Josephson measurements provide a complementary probe: when one tunneling process dominates, the neutral-channel amplitude ratio is $\Phi^{-2}$ for zero mode operator tunneling and $\Phi^{-1}$ for direct zero mode anyon transfer.

Quasiparticle poisoning can obscure these fusion-channel signatures by changing the fermion parity or topological charge stored on an FQH-SC segment \cite{rainis-2012,higginbotham-2015,albrecht-2017,karzig-2021}. Such processes allow transitions between nominally distinct fusion-channel branches and can therefore wash out the enlarged Josephson periodicity when the measurement time exceeds the poisoning time $\tau_p$. Possible mitigation strategies include improving sample quality, increasing the relevant excitation gaps, and performing branch-resolved measurements on timescales shorter than $\tau_p$. In thermal equilibrium, the defect fusion space can contribute a characteristic entropy when the temperature exceeds its residual energy splittings but remains below the bulk and proximity-induced excitation gaps. Gate-controlled changes in this entropy could be probed through charge or polarization measurements using Maxwell relations, following the proposal in Ref.~\cite{wen-2024}.

Additional tunneling processes can modify the Josephson spectrum even when quasiparticle poisoning is negligible. In Laughlin FQH-SC junctions, backscattering reduces the fractional Josephson periodicity already at zero temperature, while measurements restricted to a conserved symmetry sector can retain parafermion signatures \cite{cao-2024}. For the Read–Rezayi junction, the quoted neutral-channel amplitude ratios characterize the separate tunneling contributions; when both processes contribute appreciably, the observed response also depends on their relative microscopic amplitudes.

For multi-component, hole-conjugated, or higher-Landau-level states, the experimentally realized edge theory may depend sensitively on the confining potential and density profile \cite{wen-1995,lopez-1999}. A smooth boundary can produce an outer reconstructed strip whose topological order differs from that of the intended bulk phase. If an intervening reconstructed strip suppresses proximity coupling to the inner edge modes, superconductivity may predominantly gap the outermost modes. The resulting zero modes and tunneling signatures can then reflect the reconstructed outer phase. Controlling the edge profile and superconducting geometry is therefore essential for using zero modes to identify the underlying topological order. Interpreting the predicted tunneling signatures, therefore, requires identifying the topological order of the proximitized region and the endpoint defect types it supports.



\section*{Acknowledgments}
We thank Angela Kou for her collaboration on Ref. \cite{cao-2026-l} and during the early stages of this paper. We also thank Ramanjit Sohal for his collaboration on Ref. \cite{cao-2026-l}, his contributions during the early stages of this paper, and his helpful comments on the manuscript. We thank Erez Berg, Fiona Burnell, Paul Fendley, Hart Goldman, Eun-Ah Kim, Kyung-Su Kim, Elio K\"{o}nig, Roman Lutchyn, John McGreevy, Benjamin Moy, Seth Musser, and Pengjie Wang for helpful discussions. We especially thank Da-Chuan Lu and Taige Wang for discussions of their unpublished work. This work was supported in part by the National Science Foundation grant N. DMR 2225920 at the University of Illinois.

\appendix

\section{Symmetry and defects}\label{app:Gcross}
In this appendix, we review the elements of symmetry defect theory needed for the FQH-SC construction in the main text \cite{turaev-2000,etingof-2010,barkeshli-2019,aasen-2022}. We first introduce the $G$-crossed description of extrinsic defects, including their fusion spaces and the definition of a zero mode anyon. We then discuss the associativity transformation relevant to direct zero mode anyon transfer. For fermionic topological orders, the ordinary anyon theory is not modular because it contains a transparent physical fermion. We therefore explain how a minimal modular extension is used to incorporate fermion-parity fluxes and evaluate the Wilson and zero mode algebras \cite{gaiotto-2016,bruillard-2017,aasen-2022}. Finally, we apply these results to the Abelian, Moore–Read, and Read–Rezayi FQH-SCs considered in the main text.
\subsection{Symmetry actions and \texorpdfstring{$G$}{G}-graded defect sectors}\label{subapp:Gcross}
Let $\C$ denote the modular tensor category (MTC) describing a bosonic topological order, and let $G$ be a finite global symmetry group that acts on the anyons by invertible transformations that preserve their fusion and braiding data. In categorical language, these transformations are braided autoequivalences
\begin{align}
    \rho_{\vb{g}}:\C\ra\C\,,\quad a\mapsto\rho_{\vb{g}}(a)\,,\quad \vb{g}\in G\,.
\end{align}
We assume throughout that this action admits a consistent $G$-crossed extension; the examples considered in this work have no obstruction to introducing the corresponding symmetry defects.

The anyons and symmetry defects are organized into a $G$-graded fusion category,
\begin{align}
    \C_{G}= \bigoplus_{\vb{g}\in G} \C_{\vb{g}}\,,
\end{align}
where $\C_{\vb{1}}\equiv \C$ is the original anyon sector with $\vb{1}$ being the identity of $G$ and $\C_{\vb{g}}$ contains the defects associated with the symmetry element $\vb{g}$. An object $X_{\vb{g}}\in\C_{\vb{g}}$ specifies both the symmetry sector $\vb{g}$ and a particular defect species within that sector. The group element $\vb{g}$ therefore labels a defect sector, but it does not uniquely determine the defect species. A $\vb{g}$-defect is the endpoint of a $\vb{g}$ branch cut. When an anyon $a\in\C_{\vb{1}}$ crosses this branch cut, it is transformed according to $a\mapsto\rho_{\vb{g}}(a)$. The placement of the branch cut need not be physical; in the FQH-SC construction, it is represented by the SC segment.

For ${\vb{g}},{\vb{h}}\in G$, the defect fusion relations are analogous to those of anyons:
\begin{align}
    X_{\vb{g}}\times Y_{\vb{h}}=\sum_{Z_{\vb{gh}}\in\C_{{\vb{gh}}}}N_{X_{\vb{g}}Y_{\vb{h}}}^{Z_{\vb{gh}}} Z_{{\vb{gh}}}\,,\label{eqn:defect_f}
\end{align}
where $N_{X_{\vb{g}}Y_{\vb{h}}}^{Z_{\vb{gh}}}$ is the corresponding fusion multiplicity. The quantum dimensions satisfy
\begin{align}
    d_{X_{\vb{g}}}d_{Y_{\vb{h}}}=\sum_{Z_{\vb{gh}}\in\C_{{\vb{gh}}}}N_{X_{\vb{g}}Y_{\vb{h}}}^{Z_{\vb{gh}}} d_{Z_{{\vb{gh}}}}\,.
\end{align}
Fusion with an ordinary anyon does not change the symmetry sector: $X_{\vb{g}}\times a\in\C_{\vb{g}}$ for $a\in\C_{\vb{1}}$. It can, however, change the defect species. In the charge-conjugation defect theories considered below, the defect species within a fixed sector may be obtained by fusing a chosen reference defect with ordinary anyons. Different anyons need not generate distinct defect species, because $a\times X_{\vb{g}}\simeq a'\times X_{\vb{g}}$ may hold for topologically distinct $a$ and $a'$.

The total quantum dimension of a defect sector is 
\begin{align}
    \D_{\C_{\vb{g}}}^2\equiv \sum_{X_{\vb{g}}\in\C_{\vb{g}}}d_{X_{\vb{g}}}^2.
\end{align}
For every nonempty sector of a $G$-crossed extension, $\D_{\C_{\vb{g}}}=\D_{\C_{\vb{1}}}$, so the total quantum dimension of each defect sector equals that of the parent topological order \cite{barkeshli-2019}. 

The number of defect species is determined by the symmetry-invariant anyons. More generally, the $G$-crossed modular transformations imply that, for commuting $\vb{g},\vb{h}\in G$, the number of $\vb{h}$-invariant defects in $\C_{\vb{g}}$ equals the number of $\vb{g}$-invariant defects in $\C_{\vb{h}}$ \cite{barkeshli-2019}. With $\vb{h}=\vb{1}$, this implies that the number of topologically distinct $\vb{g}$-defect species equals the number of $\rho_{\vb{g}}$-invariant anyons in the parent topological order \cite{barkeshli-2019}:
\begin{align}
    \abs{\C_{\vb{g}}}=\abs{\C_{\vb{1}}^{\vb{g}}},\quad \C_{\vb{1}}^{\vb{g}}\equiv\cbr{a\in\C_{\vb{1}}|\rho_{\vb{g}}(a)=a}\,.\label{eqn:syminv}
\end{align}
Here $\abs{\C_{\vb{g}}}$ denotes the number of distinct simple defect species in the $\vb{g}$-sector. These relations constrain the quantum dimensions and fusion rules of the defects in the examples considered below. These invariant anyons also distinguish the defect species. If $b\in\C_{\vb{1}}$ satisfies $\rho_{\vb{g}}(b)=b$, a Wilson loop of $b$ can then be closed consistently around a $\vb{g}$-defect. Its eigenvalue depends on the defect species and provides a topological label for the defect.

\subsection{Defect fusion space and zero mode anyons}\label{subapp:defectf}
Each defect $X_{\vb{g}}$ has a conjugate defect $\bar{X}_{\vb{g}}\in\C_{\vb{g}^{-1}}$. Their fusion lies in the anyon sector, $X_{\vb{g}}\times \bar{X}_{\vb{g}}\in\C_{\vb{1}}$, and the allowed fusion outcomes determine the nonlocal Hilbert space associated with the pair,
\begin{align}
    X_{\vb{g}}\times \bar{X}_{\vb{g}}=\sum_{c\in\C_{\vb{1}}}N_{X_{\vb{g}}\bar{X}_{\vb{g}}}^c c\,.
\end{align}

The anyons $c$ appearing in this decomposition label the possible total topological charges of the defect pair. For defects at fixed positions, these fusion channels span the topologically degenerate defect Hilbert space. In the multiplicity-free examples considered below, we denote the corresponding states by $\ket{c}$.

A zero mode anyon $c_{\text{zm}}\in\C_{\vb{1}}$ is an anyon that can be absorbed or emitted by an individual defect without changing its defect species:
\begin{align}
    X_{\vb{g}}\times c_{\text{zm}}\ni X_{\vb{g}}\,,\quad N_{X_{\vb{g}}c_{\text{zm}}}^{X_{\vb{g}}}>0\,.\label{eqn:defect_zm}
\end{align}
By Frobenius reciprocity \cite{bonderson-2007-thesis,barkeshli-2019}, the fusion multiplicity in Eq.~\eqref{eqn:defect_zm} is equal to the multiplicity with which $c_{\text{zm}}$ appears in the fusion of the conjugate defect pair and the corresponding fusion multiplicity for the conjugate defect,
\begin{align}
    N_{X_{\vb{g}}c_{\text{zm}}}^{X_{\vb{g}}}=N^{c_{\text{zm}}}_{\bar{X}_{\vb{g}}X_{\vb{g}}}=N_{\bar{X}_{\vb{g}}c_{\text{zm}}}^{\bar{X}_{\vb{g}}}\,.
\end{align}
Thus, $c_{\text{zm}}$ is itself an allowed defect-pair fusion channel and a zero mode anyon for the conjugate defect. The subscript does not denote a separate type of topological excitation; it indicates that the fusion channel $c_{\text{zm}}$ additionally satisfies the defect-absorption condition in Eq.~\eqref{eqn:defect_zm}.

\subsection{\texorpdfstring{$G$}{G}-crossed associativity and direct transfer of a zero mode anyon}\label{subapp:G-F}
Associativity of anyon and defect fusion is implemented by the $G$-crossed $F$-symbols, which relate different fusion/splitting space bases \cite{bonderson-2007-thesis,barkeshli-2019}. For the process relevant to this work, consider direct transfer of a zero mode anyon $c_{\text{zm}}$ between a conjugate pair of defects $X_{\vb{g}}$ and $\bar{X}_{\vb{g}}$ \cite{bonderson-2009-split}. The two fusion bases are
\begin{align}
    \pbr{X_{\vb{g}}\times c_{\text{zm}}}\times \bar{X}_{\vb{g}}\quad\text{and}\quad X_{\vb{g}}\times \pbr{c_{\text{zm}}\times \bar{X}_{\vb{g}}}\,.
\end{align}
Because both defects can absorb $c_{\rm zm}$ without changing their species, the intermediate defect channels may remain $X_{\vb{g}}$ and $\bar{X}_{\vb{g}}$, respectively. Reassociating the fusion tree therefore gives the matrix element
\begin{align}
    T_{c_{\text{zm}}}\ket{c}=\sbr{F^{X_{\vb{g}}c_{\text{zm}}\bar{X}_{\vb{g}}}_c}_{X_{\vb{g}}\bar{X}_{\vb{g}}}\ket{c}\,,\label{eqn:zm_transfer}
\end{align}
where $c$ is the total fusion channel of the defect pair. We assume throughout that the relevant fusion spaces are multiplicity free; otherwise, Eq.~\eqref{eqn:zm_transfer} is a matrix acting on the corresponding multiplicity indices.

An individual $F$-symbol depends on the choice of bases for the fusion and splitting spaces. Under a fusion-vertex gauge transformation, the matrix element in Eq.~\eqref{eqn:zm_transfer} is accompanied by the opposite transformation of the microscopic tunneling coefficient, so the complete physical tunneling amplitude is gauge invariant. In the multiplicity-free examples considered below, the relative matrix elements between different fusion channels are gauge invariant, while their common overall phase is absorbed into the microscopic tunneling amplitude.

\subsection{Fermionic topological orders and minimal modular extensions}\label{subapp:mme}
Electronic FQH states contain a transparent physical fermion $\Psi$, satisfying
\begin{align}
    \Psi\times\Psi=0\,,\quad M_{a\Psi}=1\,,\quad \forall a\in\C\,.
\end{align}
Because $\Psi$ has trivial monodromy with every physical anyon, the corresponding braided fusion category $\B$---which relates to $\C$ by $\C\equiv \B/{\rm sVec}$---has a transparent subcategory ${\rm sVec}\equiv\cbr{0,\Psi}$. Its $\S$-matrix is degenerate, so $\B$ is therefore super-modular rather than modular. In particular, the physical sectors $a$ and $a\times\Psi$ cannot be distinguished by Wilson loops constructed entirely from physical anyons. 

This problem is resolved by embedding $\B$ into a minimal modular extension $\M$ \cite{gaiotto-2016,bruillard-2017,aasen-2022}. The extension contains the physical anyon theory as a full braided subcategory and can be decomposed as
\begin{align}
    \M\equiv \M_+\oplus\M_-\,,\quad \M_+\equiv \B\,.
\end{align}
The sector $\M_+$ contains the physical anyons, including the transparent fermion $\Psi$, while $\M_-$ contains fermion-parity fluxes, namely excitations with nontrivial monodromy with $\Psi$. The extension is called minimal when it introduces only the sectors required to lift the degeneracy of the braiding data. Equivalently, its total quantum dimension satisfies
\begin{align}
    \D_{\M}^2=2\D_{\B}^2=2\sum_{a\in\B}d_a^2\,.
\end{align}

The fermion-parity fluxes in $\M_-$ are not additional deconfined quasiparticles of the original electronic FQH state. Rather, $\M$ is an auxiliary modular theory obtained by gauging fermion parity \cite{bruillard-2017,gaiotto-2016}. Its nondegenerate modular data allow us to evaluate the Wilson algebra and construct the corresponding charge-conjugation defect theory. Nevertheless, the parity-flux sectors can contribute to the FQH-SC defect Hilbert space by distinguishing states that differ in nonlocal fermion parity. This accounts for the additional fermion-parity contribution to the zero mode structure of electronic FQH-SCs \cite{kitaev-2001,barkeshli-2013,aasen-2022}.

For the minimal fermionic theory ($\text{sVec}$), the possible minimal modular extensions form Kitaev's 16-fold way \cite{kitaev-2006,bruillard-2017,aasen-2022}. They are labeled by an integer $\nu_{16}\in \Z_{16}$, related to the chiral central charge $c_f$ \footnote{When $\B=\C\boxtimes{\rm sVec}$ with $\C$ modular, the central charges satisfy $c_-(\rm FQH)=c_-(\C)+c_f\pmod 8$, where $c_f$ is the chiral central charge of the chosen minimal modular extension of ${\rm sVec}$.} of the chosen minimal modular extension of ${\rm sVec}$ by 
\begin{align}
    c_f=\frac{\nu_{16}}{2}\text{ mod 8}.
\end{align}
When $\nu_{16}$ is odd, the fermion-parity-flux sector contains a non-Abelian Ising-like flux. When $\nu_{16}$ is even, the extension is Abelian. In the examples below, we specify only the modular extension selected by the chiral central charge and anyon content of the particular FQH state. When the extension factorizes, the appropriate 16-fold-way sector is combined with these additional sectors. For example, the fermionic Laughlin state is conveniently embedded into $U(1)_{4m}$, while the generalized Moore–Read and Read–Rezayi states admit the modular extensions specified in their respective subsections. 

The chiral central charge can be determined \footnote{We thank Seth Musser for pointing us to Ref. \cite{lapa-2019}.} modulo eight by the charge-dependent Gauss–Milgram relation for fermionic theories with charge conservation \cite{lapa-2019}
\begin{align}
    e^{\frac{2\pi i}{8}(c_--\si_{H})}=\frac{1}{\sqrt{2}\mathcal D_{\B}}\sum_{a\in\mathcal B} d_a^2\th_a e^{i\pi q_a/e}\,,
\end{align}
where $\sigma_H$ is the Hall conductivity in units of $e^2/h$, and $q_a$ is the electric charge of anyon $a$ modulo $2e$, and $\th_a$ is the topological spin. The sum runs over the physical sectors of the super-modular category $\mathcal B$, with the electron retained as a distinct object. Given $\sigma_H$, this relation determines $c_-$ modulo $8$ without explicitly introducing the fermion-parity flux sectors.

The fermion-parity fluxes introduced by the minimal modular extension correspond to primary fields with monodromy $-1$ with the charge-$e$ electron simple current, $\Psi_\pm$ \cite{moore-1991}. An electron acquires a minus sign upon encircling such a field, so these primary fields act as twist fields for the electron and belong to the Ramond sectors of the fermionic chiral algebra \cite{cheng-2026-order}. They nevertheless have trivial monodromy with the charge-$2e$ simple current, $J_{2e,\pm}=\Psi_\pm\times\Psi_\pm$. In our construction, the CFT corresponding to the minimal modular extension is obtained by extending the product of the neutral and charge chiral algebras by $J_{2e,\pm}$ \cite{bhardwaj-2017}. This extension retains both the electron-local primary fields and the fermion-parity-flux primary fields, distinguished by their monodromy $+1$ and $-1$ with $\Psi_\pm$, respectively.

\subsection{Charge-conjugation defect calculations}
The preceding results provide a common procedure for determining the charge-conjugation defects and zero mode structure of the examples below. For a bosonic topological order, the calculation is performed directly in its MTC $\C$. For a fermionic FQH state, we instead work in a minimal modular extension $\M$, as described in the previous subsection.

We first specify the action of charge conjugation $\rho_{\vb{C}}(a)=\bar{a}$ and identify the charge-conjugation-invariant anyons satisfying $\rho_{\vb{C}}(a)=a$. Equation \eqref{eqn:syminv} then determines the number of distinct charge-conjugation defect species. We choose a reference defect $X_0$ and generate the remaining species by fusing $X_0$ with ordinary anyons, identifying anyon dressings that produce topologically equivalent defects.

The invariant-anyon counting fixes the number of defect species but does not by itself determine their fusion rules. To obtain the latter, we use the $G$-grading, compatibility with fusion by ordinary anyons, and total quantum dimension $\D_{\C_{\vb{C}}}=\D_{\C_{\vb{1}}}$. Wilson loops of charge-conjugation-invariant anyons may then be used to distinguish the resulting defect species.

For each defect $X_{\vb{C}}$, the zero mode anyons are identified using Eq.~\eqref{eqn:Gzm_def}. By Frobenius reciprocity, the same $c_{\text{zm}}$ appears as a fusion channel of the corresponding conjugate defect pair. Equation \eqref{eqn:zm_def} then determines which terminating Wilson-line labels $a$ provide localized operator realizations of these zero mode anyons. For fermionic FQH states, these steps are carried out in the minimal modular extension. Contributions associated with fermion-parity fluxes are retained when they encode a physical nonlocal fermion-parity degree of freedom of the FQH-SC.

\subsubsection{Bosonic, Abelian FQH-SC with \texorpdfstring{$\Z_m$}{Zm} anyons}\label{subapp:bZm}
We first consider a bosonic Abelian FQH state whose anyons are labeled by $j\in\Z_m$, with fusion $j\times k=(j+k)\text{ mod } m$. Charge conjugation acts as $\rho_{\vb{C}}(j)=-j\text{ mod } m$. The defect structure depends on whether $m$ is even or odd. For even $m$, the charge-conjugation-invariant anyons are $0$ and $m/2$, so the charge-conjugation sector contains two defect species $X_\pm$. Their quantum dimensions are $d_{X_\pm}=\sqrt{m/2}$ and the fusion rules are \cite{barkeshli-2019}
\begin{align}
    X_\pm\times \bar{X}_\pm=\sum_{k\in\Z_{m/2}}2k\,&,\quad X_+\times \bar{X}_-=\sum_{k\in\Z_{m/2}} (2k+1)\,,\\
    X_\pm\times 2k= X_\pm\,&,\quad X_\pm\times (2k+1)=X_\mp\,.
\end{align}
The zero mode anyons form the even subgroup $2\Z_{m}\simeq\Z_{m/2}$, giving a $\Z_{m/2}$ zero mode structure.

For odd $m$ (for example, the Halperin $(2,2,1)$ state which corresponds to $m=3$), there is only one invariant anyon $j=0$, implying that there is a single charge-conjugation defect that has the following fusion rules and quantum dimension,
\begin{align}
    X\times X=\sum_{j\in \Z_m} j\,,\quad X\times j=X\,,\quad d_X=\sqrt{m}.
\end{align}
Every anyon is therefore a zero mode anyon, giving a $\Z_m$ zero mode structure.
\subsubsection{Fermionic, Abelian FQH-SC with \texorpdfstring{$\Z_m$}{Zm} anyons}\label{subsubsec:fAb}
We next consider the class of fermionic Abelian FQH states whose super-modular category factorizes as $\M_+=\{0,\psi\}\boxtimes \Z_m$. For an Abelian modular theory with anyon group $\A$ and charge-conjugation action, the zero mode anyons supported by a fixed charge-conjugation defect form the doubled subgroup $2\A$. The appropriate minimal modular extension depends on the chiral central charge. When $c_f$ is odd, $\nu_{16}=2\text{ mod }4$, and the fermion-parity sector has $\Z_4$-type fusion rules. The modular extension may therefore be written as $\M=\Z_4\boxtimes \Z_m$, and the zero mode anyons form the following subgroups
\begin{equation}
    2\A_\M=2\Z_4\times 2\Z_m\simeq\begin{cases}
        \Z_2\times\Z_m\simeq \Z_{2m} &\quad\text{for odd }m,\\
        \Z_2\times\Z_{m/2} &\quad\text{for even }m,
    \end{cases}\label{eqn:fzm}
\end{equation}
The additional $\Z_2$ factor originates from the fermion-parity-flux sector and represents the nonlocal fermion-parity degree of freedom of the FQH-SC.

When $c_f$ is even, $\nu_{16}=0\text{ mod }4$, and the fermion-parity sector has toric-code-like $\Z_2\times\Z_2$ fusion rules. Since doubling annihilates both order-two factors, they do not contribute an additional zero mode subgroup. The resulting structure is therefore 
\begin{equation}
    2\A_\M\simeq\begin{cases}
        \Z_{m} &\quad\text{for odd }m,\\
        \Z_{m/2} &\quad\text{for even }m,
    \end{cases}
\end{equation}
The fermionic Laughlin state at filling fraction $\nu=1/m$ has odd $m$ and $c_f$ is odd. Its minimal modular extension is $U(1)_{4m}$, equivalently the cyclic realization of $\Z_4\boxtimes\Z_m$, and Eq.~\eqref{eqn:fzm} gives the familiar $\Z_{2m}$ zero mode structure.
\subsubsection{Bosonic Moore-Read FQH-SC at \texorpdfstring{$\nu=1/m$}{1/m}}\label{subapp:bMRdefect}
For the bosonic generalized Moore–Read state at filling fraction $\nu=1/m$, with $m$ odd, the parent topological order consists of the Ising neutral sector and the $U(1)_{4m}$ charge sector. We label the anyons by $a_l$, where $a\in\{1,\si,\chi\}$ and $l\in\Z_{4m}$, subject to locality with respect to the physical boson
\begin{align}
    2j_a+l\in2\Z\,,\quad a_l\sim (a\times\chi)_{l+2m}\,,
\end{align}
with $j_1=0,\;j_\si=1/2,\;j_\chi=1$. The local physical boson is $\chi_{2m}$. The charge-conjugation-invariant anyons are $1_0$, $\chi_0$ and $\si_{m}$, so Eq.~\eqref{eqn:syminv} implies that there are three charge-conjugation defect species, which we denote by $X_1$, $X_\chi$, $X_\si$. Their fusion rules can be written compactly as
\begin{align}
    X_a=X_1\times a_l\,,\quad X_a\times X_b =\sum_{\substack{c\in\{1,\si,\chi\}\\l\in\Z_2}}\sum_{\substack{k\in \Z_{m}\\2j_c+l\in 2\Z}} N_{ab}^c\,c_{l+4k}\,,
\end{align}
with the condition $c_l\sim (c\times \chi)_{l+2m}$. All three defect species support the charged zero mode anyons $1_{4k}$, which form a $\Z_{m}$ sector generated by $1_4$. The defect $X_\si$ supports the neutral zero mode anyon $\chi_0$. Since $\chi_0\times1_{4k}=\chi_{4k}$, the zero mode structure can be organized as $\Z_2^{(n)}\times \Z_m$, where $\Z_2^{(n)}\equiv\cbr{1_0,\chi_0}$ is the neutral Ising fusion sector.
\subsubsection{Fermionic Moore-Read FQH-SC at \texorpdfstring{$\nu=1/m$}{1/m}}\label{subapp:fMRdefect}
For fermionic MR FQH-SC at filling $\nu=1/m$, with $m$ even, we work in the modular extension $\M=\text{Ising}\,\boxtimes\,U(1)_{4m}$ whose anyons are labeled $a_l$ with $l\in\Z_{4m}$. The total quantum dimension is $\D_{\M}=\sqrt{16m}$. The charge-conjugation-invariant anyons are $1_0$, $1_{2m}$, $\chi_0$, $\chi_{2m}$, $\si_0$, and $\si_{2m}$. Equation \eqref{eqn:syminv} implies that the charge-conjugation sector contains six defect species, which we label by $X_{a_{l\text{ mod } 2}}$. As in the bosonic case, there is a bare defect $X_{1_0}$ and the other defects are related by
\begin{align}
    X_{a_{l\text{ mod } 2}}\equiv X_{1_0}\times a_l\,.\label{eqn:fMR_def+anyon}
\end{align}
Their fusion rules can be written compactly as
\begin{align}
    X_{a_l}\times X_{b_k}=\sum_{\substack{c\in\{1,\si,\chi\}\\n\in \Z_{2m}}}N_{ab}^c~ c_{(l+k)\text{ mod 2}+2n}\,.
\end{align}
Similar to the bosonic case, all defects support the $\Z_{2m}$ zero mode anyons generated by $1_2$, and the defects $X_{\si_0}$ and $X_{\si_1}$ support an additional neutral zero mode anyon $\chi_0$, making the zero mode structure $\Z_2^{(n)}\times\Z_{2m}$. 
\subsubsection{Bosonic Read-Rezayi FQH-SC at \texorpdfstring{$\nu=3/(3M+2)$}{3/(3M+2)}}\label{subapp:bRRdefect}
For bosonic RR FQH-SC at filling $\nu=3/(3M+2)$, with $M$ even, the parent topological order has chiral algebra \cite{seiberg-2016}
\begin{align}
    \frac{{\rm Para}_3\times U(1)_{3(3M+2)}}{\Z_3}\,,\quad {\rm Para}_3\equiv \frac{SU(2)_3}{U(1)_6}\,.
\end{align}
By matching the topological spin of primary fields and anyons in the charge sectors, the anyons in RR FQH can be reorganized as ${\rm Fib}\boxtimes \Z_{3M+2}^{(2M+3/2)}$. The anyons invariant under charge-conjugation are $1_0$, $1_{\frac
{3M+2}{2}}$, $\tau_0$ and $\tau_{\frac
{3M+2}{2}}$. Equation \eqref{eqn:syminv} implies that the charge-conjugation sector contains four defect species. We label the bare defect by $X_{1_0}$ and other defects by 
\begin{align}
    X_{a_{l\text{ mod } 2}}\equiv X_{1_0}\times a_l\,.\label{eqn:bRR_def+anyon}
\end{align}
Their fusion rules can be written compactly as
\begin{align}
    X_{a_l}\times X_{b_k}=\sum_{\substack{c\in\{1,\tau\}\\n\in\Z_{\frac{3M+2}{2}}}} N_{ab}^c~ c_{(l+k)\text{ mod 2}+2n}\,,\quad X_{a_l}\times b_j=\sum_{c\in\{1,\tau\}} N_{ab}^c\,X_{c_{(l+j)\text{ mod 2}}}\,,
\end{align}
where $j\in\Z_{3M+2}$ labels the charge sector of an anyon in RR FQH. All defect species support the charged zero mode anyons $1_{2k}$, which form a $\Z_{\frac{3M+2}{2}}$ sector generated by $1_2$. The defects $X_{\tau_0}$ and $X_{\tau_1}$ support the neutral zero mode anyon $\tau_0$. Their zero mode structure can be organized as ${\rm Fib}\times \Z_{\frac{3M+2}{2}}$.

\subsubsection{Fermionic Read-Rezayi FQH-SC at \texorpdfstring{$\nu=3/(3M+2)$}{3/3(M+2)}}\label{subapp:RRdefect}
For RR FQH-SC at filling $\nu=3/(3M+2)$, with $M$ odd, the anyons can be reorganized as ${\rm Fib}\boxtimes \Z_{3M+2}^{((M+1)/2)}$. The RR FQH has chiral central charge $c_-=\frac{9}{5}$ \cite{read-1999} while ${\rm Fib}$ has chiral central charge $c_-=\frac{14}{5}$. The charge sector of the minimal modular extension must therefore have chiral central charge $-1\,{\rm mod }\,8$. A compatible minimal modular extension is $\M=\text{Fib }\boxtimes\Z_{4(3M+2)}^{(3/2)}$. The charge-sector defect fusion rules therefore give the same $\Z_{2(3M+2)}$ zero mode structure as in the fermionic Abelian case. The $X_{\tau_0}$ and $X_{\tau_1}$ species additionally absorb $\tau_0$, giving the ${\rm Fib}\times\Z_{2(3M+2)}$ structure.
\section{RG analysis of FQH-SC theory}\label{app:rg}
In this appendix, we determine the beta functions of the superconducting pairing terms used in the main text. Density-density interactions between the counter-propagating edges renormalize both the charge-mode velocity $u$ and Luttinger parameter $K$, thereby modifying the scaling dimension of the pairing perturbation. For each example, we derive the corresponding nonchiral boundary theory and evaluate the scaling dimension of the Cooper-pairing operator at the unpaired fixed point $\De=0$. These conditions specify when the fully gapped FQH-SC phase assumed in Sec.~\ref{sec:setup} can be realized. Throughout, $\De$ denotes the dimensionless pairing amplitude.
\subsection{Laughlin FQH-SC}\label{subapp:Lrg}
For Laughlin FQH-SC, we follow Ref. \cite{cao-2024}. The counter-propagating edge theory is $\L_{U(1)_{4m},+}+\L_{U(1)_{4m},-}$. We include a Cooper pairing term $\De\Psi_+\Psi_-+\hc\sim\De \cos\sbr{2m(\phi_+-\phi_-)}$ and a density-density interaction term $-\frac{4mv_FU}{2\pi}\pa_x\phi_+\pa_x\phi_-$. By defining nonchiral fields $\vp,\vt=(\phi_+\pm\phi_-)/2$, the Lagrangian takes the sine-Gordon form \cite{schiller-2023},
\begin{align}
    L_{1/m,\De}=\int_{0}^L\rd x\,\cbr{\frac{4m}{2\pi}\frac{1}{K}\sbr{\frac{1}{u}\pbr{\pa_t\vt}^2-u\pbr{\pa_x\vt}^2}+\De\cos(4m\vt)}\,,
\end{align}
where we leave the spacetime coordinates of $\vt$ implicit for brevity. The density-density interaction renormalizes the velocity $u$ and the Luttinger parameter $K$ to be,
\begin{align}
    u=v_F\sqrt{1-U^2}\,,\quad K=\sqrt{\frac{1+U}{1-U}}\,,
\end{align}
where $v_F$ is the Fermi velocity of uncoupled edges. The pairing operator $\cos(4m\vt)$ has scaling dimension $mK$. Defining $x\equiv 2-mK$ which measures the distance from marginality, the leading RG equations near $(x,\De)=(0,0)$ take the Kosterlitz form \cite{kosterlitz-1974,amit-1980}:
\begin{align}
    \be(\De)=x\De+...\,,\quad\be(x)=32\pi^3\De^2+...\,.
\end{align}
The superconducting perturbation is relevant at weak coupling for $K<2/m$.
\subsection{Moore-Read FQH-SC}\label{subapp:MRrg}
The MR FQH-SC is described by a Cooper pairing term $\De(\Psi_+\Psi_-+\hc)\sim i\De\chi_+\chi_-$ $\cos[2m(\phi_+-\phi_-)]$ and a density-density interaction, resulting in the following Lagrangian,
\begin{align}
    \L_{\text{MR},\De}&=\L_{\text{MR},+}+\L_{\text{MR},-}-\frac{4mv_c U}{2\pi}\pa_x\phi_+\pa_x\phi_-+i\De \chi_+\chi_-\cos[2m(\phi_+-\phi_-)]\,,
\end{align}
with units where the velocity of the neutral mode is set to $v_n=1$. We can write this theory in terms of a nonchiral Majorana fermion $\chi$ and a nonchiral boson $\vt$,
\begin{align}
    \L_{\text{MR},\De}=\frac{i}{2}\bar{\chi}\sl{\pa}\chi+\frac{4m}{2\pi}\frac{1}{K}\sbr{\frac{1}{u}\pbr{\pa_t\vt}^2-u\pbr{\pa_x\vt}^2}-\frac{\De}{2}\bar{\chi}\chi\cos(4m\vt)\,.
\end{align}
The density-density interaction renormalizes the velocity of the charge mode $u$ and the Luttinger parameter $K$ to be
\begin{align}
    u=v_c\sqrt{1-U^2},\quad K=\sqrt{\frac{1+U}{1-U}}\,.
\end{align}
The scaling dimensions of the vertex operator and Majorana fermion follow from two-point correlation functions at the MR CFT fixed point,
\begin{align}
    \ev{e^{il\vt(0)}e^{-il\vt(\vec{x})}}&\propto\frac{1}{\abs{\vec{x}}^{2\De_l}}\,,\quad \De_l=\frac{l^2K}{16m}\,,\\
    \abs{\ev{\chi_\pm(0)\chi_\pm(\vec{x})}}&\propto\frac{1}{\abs{\vec{x}}^{2\De_\chi}}\,, \quad\De_\chi=\frac{1}{2}\,,
\end{align}
where $\vec{x}$ denotes the separation in the (1+1)-dimensional spacetime. This means that the Cooper pairing operator has scaling dimension $\De_{4m}+2\De_\chi=mK+1$. Thus, for $m>1$, the superconducting perturbation is irrelevant without density-density interactions, where $K=1$, and becomes relevant only for sufficiently attractive interactions such that $K<1/m$. For $m=1$, it is marginal at tree level when $K=1$. Near the critical point $K_c=1/m$, we define $x\equiv1-mK$. The one-loop RG equations are
\begin{align}
    \be(\De)=x\De+...\,,\quad\be(x)=4\pi^3\De^2+...\,.
\end{align}
These RG equations have the same Kosterlitz RG structure as the Abelian case. The agreement with the Abelian Kosterlitz flow is expected because Eq.~\eqref{eqn:MRD} lies in the same universality class as the supersymmetric sine-Gordon model \cite{witten-1978} which is known to admit Kosterlitz RG \cite{goldschmidt-1986}.

\subsection{Read-Rezayi FQH-SC}\label{subapp:RRrg}
The RR FQH-SC is described by the pairing perturbation $\De\Psi_+\Psi_-+\hc=\De \psi_{1,+}\psi_{1,-}$ $e^{4(3M+2)i\vt}+\hc$, and a density-density interaction in the charge sector. In terms of the nonchiral field $\vt$ defined above, the resulting Lagrangian is
\begin{align*}
    \L_{\text{RR},\De}=\L_{\Z_3,+}+\L_{\Z_3,-}+\frac{12(3M+2)}{2\pi}\frac{1}{K}\sbr{\frac{1}{u}(\pa_t\vt)^2-u(\pa_x\vt)^2}+\De\psi_{1,+}\psi_{1,-}e^{4(3M+2)i\vt}+\hc\,.
\end{align*}
With this normalization, the charge-sector vertex operator has scaling dimension $\De_l=\frac{l^2 K}{48(3M+2)}$. Since the chiral $\Z_3$ parafermion current has conformal dimension $h_{\psi_1}=2/3$, the full pairing operator has scaling dimension
\begin{align}
    \De_{\text{SC}}=\frac{4}{3}+\frac{(3M+2)K}{3}\,.
\end{align}
Defining $x\equiv2-\De_{\text{SC}}=[2-(3M+2)K]/3$, the RG equations admit the same Kosterlitz form,
\begin{align}
    \be(\De)=x\De+...\,,\quad \be(x)=\frac{64\pi^3}{9}\De^2+...\,.
\end{align}
The superconducting perturbation is therefore relevant at weak coupling for $K<2/(3M+2)$. Thus, sufficiently attractive charge-sector interactions can drive the pairing perturbation toward the strong-coupling FQH-SC phase assumed in Sec.~\ref{subsec:RR}.

\section{\texorpdfstring{$K$}{K}-matrix Formalism and General Abelian FQH-SCs}\label{app:abelian}
\subsection{Abelian FQH states and the \texorpdfstring{$K$}{K}-matrix formalism}\label{app:kmat}

In this section, we review the $K$-matrix theory for multicomponent FQH states \cite{wen-1992,wen-1995,lopez-1999,wen-book-2004}. An $N$-component FQH state is specified by a symmetric, nondegenerate, integer-valued $N\times N$ matrix $K_{IJ}$ and an integer-valued charge vector $t_I$:
\begin{align}
    \L_{K}=\frac{K_{IJ}}{4\pi}a^I\wedge da^J+\frac{1}{2\pi}t_I A\wedge da^I+j_{qp}l_Ia^I\,,\label{eqn:K}
\end{align}
where $\wedge$ is the wedge product that we leave implicit for brevity in the rest of this section, $I,J=1,...,N$ are the indices for the $K$-matrix (and we assume repeated up and down indices are summed over), $A$ is a background $U(1)$ gauge field, $j_{qp}$ describes the quasiparticle currents and $l_I$ are used to label quasiparticles. A bosonic $K$-matrix has even diagonal entries, whereas a fermionic $K$-matrix has at least one odd diagonal entry. On an open manifold, the boundary theory can be described by $N$ left/right-moving chiral bosons
\begin{align}
    \L_{K\text{,bdy}}=-\frac{K_{IJ}}{4\pi}\pa_t\phi^I\pa_x\phi^J-\frac{V_{IJ}}{4\pi}\pa_x\phi^I\pa_x\phi^J\,,\label{eqn:Kbdy}
\end{align}
where $V_{IJ}$ is a positive-definite matrix that describes the strength of density-density interaction. We orient the boundary coordinate so that positive and negative eigenvalues of $K$ correspond to right- and left-moving modes, respectively. The numbers of chiral and anti-chiral edge modes are determined by the signature of $K$. Canonical quantization gives
\begin{align}
    \comm{\phi^I(x_1)}{\pa_{x_2}\phi^J(x_2)}=-2\pi i \pbr{K^{-1}}^{IJ}\de(x_1-x_2)\,,
\end{align}
where $\pbr{K^{-1}}^{IJ}$ is the inverse of $K_{IJ}$. Different $K$-matrix theories are topologically equivalent when they are related by an integer basis transformation $W\in GL(N,\Z)$,
\begin{align}
    \pbr{K'}_{LM}=W_L^{\phantom{L}I}K_{IJ} \pbr{W^T}^J_{\phantom{J}M}\,,\quad \pbr{t'}_J=W_J^{\phantom{J}I}t_I\,,
\end{align}
where $W^T$ is the transpose of $W$ and $\det W=\pm1$.

The filling fraction and the ground state degeneracy on a genus-$g$ closed surface are
\begin{align}
    \nu=t_I\pbr{K^{-1}}^{IJ}t_J\,,\quad {\rm GSD}_g=\abs{\det K}^g\,.
\end{align}
Quasiparticles are labeled by integer vectors $l$ that correspond to vertex operators $\psi_{qp,l}=\exp(il_I\phi^I)$. Their electric charge and mutual monodromy are
\begin{align}
    q_j=et_I\pbr{K^{-1}}^{IJ}l_{j,J},\quad M_{jk}=\exp\sbr{2\pi i l_{j,I}\pbr{K^{-1}}^{IJ}l_{k,J}}\,.
\end{align}
Local excitations are labeled by vectors $l_{I,\text{local}}=K_{IJ}\La^J$ where $\La^J\in\Z$. A physical electron is a local excitation with electric charge $e$ and fermionic self-statistics. In the examples below, the rows of $K$ provide convenient choices of electron vectors.

A quasiparticle worldline along a closed curve $\ga$ is represented by the Wilson operator
\begin{align}
    W_{\ga,j}=\exp(i\oint_\ga l_{j,I}a^I)\,.
\end{align}
An open Wilson line with endpoints $x_1$ and $x_2$ on the boundary creates a quasiparticle at one endpoint, transports it through the bulk, and annihilates it at the other:
\begin{align}
    W_{(x_1,x_2),j}=\exp\cbr{il_{j,I}\sbr{\phi_I(x_1)-\phi_I(x_2)}} \exp(i\int_{x_1}^{x_2} l_{j,I}a^I)\,.
\end{align}
\subsection{\texorpdfstring{$\nu=2/5$}{2/5} FQH states: polarized Jain and \texorpdfstring{$(332)$}{(332)} Halperin state}
There are two dominant FQH states at filling fraction $\nu=2/5$, the spin-polarized Jain sequence state \cite{haldane-1983,halperin-1983,halperin-1984,jain-1989a,jain-1989b,jain-1990} and the $(332)$ Halperin spin-singlet state \cite{halperin-1983,halperin-1984}. Both states are described by the following $K$-matrix \cite{wen-1992,wen-1995,wen-book-2004},
\begin{align}
    K=\mqty(3&2\\2&3)\,,\quad t=\mqty(1\\1)\,.
\end{align}
The anyon group is $\A_K=\Z^2/K\Z^2\simeq \Z_5$ and its braided data correspond to $\Z_5^{(4)}$. We consider the minimal modular extension $\Z_5^{(4)}\boxtimes U(1)_2^2$. The two $U(1)_2$ factors do not supply an additional zero mode factor in this case because all of their anyons square to the vacuum. The resulting intrinsic zero mode structure is therefore $\Z_5$.

Although the two states share the same intrinsic Abelian topological order, they differ in their symmetry and geometric-response data. These distinctions can be encoded through their physical-spin assignments and the Wen–Zee orbital-spin vector \cite{wen-1992-spin,lee-1994,einarsson-1995}. In the planar geometry considered here, the most relevant difference is the polarization of the electronic degrees of freedom. The Jain sequence state is spin polarized, whereas the Halperin $(332)$ state is a spin singlet with an additional $SU(2)$ symmetry \cite{balatsky-1991}. This additional symmetry constrains the superconducting gapping terms allowed at the boundary \footnote{This additional $SU(2)$ symmetry will not affect symmetry fractionalization of the FQH state as $H^2\big(U(1)_c\times SU(2),\A\big)=H^2(U(1)_c,\A)=\A$.}.

The anyons in the theory are labeled by $l_I=(l_1,l_2)$ and are represented by the edge vertex operators $e^{i(l_1\phi_1+l_2\phi_2)}$. The electron operators are $\Psi_1\equiv e^{i(3\phi_1+2\phi_2)}$ and $\Psi_2\equiv e^{i(2\phi_1+3\phi_2)}$. In the polarized Jain state, $\Psi_1$ and $\Psi_2$ are two local representatives of the same spin-polarized electronic sector. In the Halperin spin-singlet state, they instead carry opposite spin and form an SU(2) doublet. This distinction changes the symmetry-allowed Cooper-pairing perturbations and the local representatives of the zero mode operators, but not the intrinsic $\Z_5$ defect structure.

For the polarized Jain state, the symmetry-allowed pairing perturbation between the two counter-propagating edges is
\begin{align}
    \Psi_{1,+}\Psi_{1,-}+\Psi_{2,+}\Psi_{2,-}+\hc=2\sbr{\cos(6\vt_1+4\vt_2-\frac{3\pi}{2})+\cos(4\vt_1+6\vt_2-\frac{3\pi}{2})}\,,
\end{align}
where $\vp_{I},\vt_{I}\equiv (\phi_{I,+}\pm\phi_{I,-})/2$. The phase shifts arise from the commutation relations between bosonic fields belonging to different components. The zero mode operators may be represented by
\begin{align}
    \al_{(1,0)}\approx e^{i2\vt_1}\sim\al_{(0,1)}\approx e^{i2\vt_2}\,.\label{eqn:Jainzm}
\end{align}
The corresponding projected symmetry operator acts as
\begin{align}
    U_{(1,1)}=e^{2i(\vp_1+\vp_2)}\,,\quad U_{(1,1)}:\phi_{I,\pm}\ra\phi_{I,\pm}\mp \frac{2\pi}{5}\,,\quad \vt_I\ra\vt_I-\frac{2\pi}{5}\,.
\end{align}
These operators satisfy the $\Z_5$ PZM algebra. Moreover, 
\begin{align}
    U_{(1,1)}^\da \Psi_{I,\pm} U_{(1,1)}=\Psi_{I,\pm}\,.
\end{align}
The operator $U_{(1,1)}$ acts nontrivially on the $\Z_5$ zero mode sector while leaving local electron operators invariant, reflecting the fact that fermion parity does not contribute to the defect Hilbert space.

For the Halperin $(332)$ state, the two electron operators carry opposite spin. The pairing can be decomposed into spin-singlet and spin-triplet channels. A nonzero spin-triplet pairing amplitude breaks the $SU(2)$ spin symmetry.
\begin{align}
    \de^{IJ}\Psi_{I,+}\Psi_{J,-}+\hc=2\sbr{\cos(6\vt_1+4\vt_2-\frac{3\pi}{2})+\cos(4\vt_1+6\vt_2-\frac{3\pi}{2})}\,.
\end{align}
The spin-singlet pairing perturbation instead takes the form
\begin{align}
    \ep^{IJ}\Psi_{I,+}\Psi_{J,-}+\hc=2[\cos(5\vt_1+\vp_1+5\vt_2-\vp_2-\pi)-\cos(5\vt_1-\vp_1+5\vt_2+\vp_2-\pi)]\,.
\end{align}

For triplet pairing, the local zero mode representatives are again those in Eq.~\eqref{eqn:Jainzm}. The two operators now carry distinct spin quantum numbers, although this distinction does not change the intrinsic $\Z_5$ defect fusion space. For singlet pairing, convenient local representatives are
\begin{align}
    \al_{(1,0)}\approx e^{i(\vt_1+\vp_1+\vt_2-\vp_2)}\,,\quad\al_{(0,1)}\approx e^{i(\vt_1-\vp_1+\vt_2+\vp_2)}\,.
\end{align}
The projected ground-state operator remains $U_{(1,1)}$, because it descends from the same bulk Wilson loop and is independent of the choice of superconducting pairing channel. Its algebra with the zero mode operators again gives the $\Z_5$ parafermion structure.

Thus, the polarized Jain and Halperin $(332)$ FQH-SCs share the same intrinsic $\Z_5$ zero mode structure and $10\pi$ Josephson periodicity. Their distinction appears in the spin quantum numbers and local edge-theory representatives of the zero mode operators, rather than in the defect fusion space.
\subsection{Hole-conjugated states at filling fraction \texorpdfstring{$\nu=2/3$}{2/3} and \texorpdfstring{$\nu=3/5$}{3/5}}\label{app:hole}
Compared with their particle-like partners, hole-conjugated FQH states contain an additional counterpropagating edge mode. This mode changes the chiral central charge and hence the fermionic modular extension, while the pairing terms in the fractional sector remain closely related to those of the corresponding particle-like states \footnote{Pairing within the same edge $\C_+$ requires appropriate momentum matching or a mechanism that supplies the missing momentum. We do not consider such pairing here.}.

At filling fraction $\nu=2/3$, the spin-polarized Jain state and the Halperin $(1,1,2)$ spin-singlet state are both described by
\begin{align}
    K=\mqty(1&2\\2&1),\quad t=\mqty(1\\1)\,.
\end{align}
The corresponding anyon theory is $\Z_3^{(1)}$, with minimal modular extension $\Z_3^{(1)}\boxtimes U(1)_{-2}^2$. Because the $U(1)_{-2}$ factors do not contribute an additional zero mode sector, the resulting FQH-SC has a $\Z_3$ PZM structure. For the polarized Jain state, or for a triplet-paired Halperin state, the same-species pairing perturbation is
\begin{align}
    \de_{IJ}\Psi_{I,+}\Psi_{J,-}+\hc=2\sbr{\cos(2\vt_1+4\vt_2-\frac{\pi}{2})+\cos(4\vt_1+2\vt_2-\frac{\pi}{2})}\,,\notag
\end{align}
with the zero mode operators and symmetry operator:
\begin{align}
    \al_{(1,0)}\approx e^{2i\vt_1}\,,\quad\al_{(0,1)}\approx e^{2i\vt_2}\,,\quad U_{(1,1)}=e^{2i(\vp_1+\vp_2)}\,.
\end{align}
These operators satisfy the $\Z_3$ PZM algebra.

At filling fraction $\nu=3/5$, the hole-conjugated Jain and Halperin states are described by
\begin{align}
    K=\mqty(-3&-2&0\\-2&-3&0\\0&0&1)\,,\quad t=\mqty(1\\1\\1)\,.
\end{align}
The physical anyon group is $\Z_5^{(1)}$. Because the state has odd chiral central charge, its minimal modular extension has fusion group $\Z_4\times\Z_5$ with zero mode structure $\Z_2\times\Z_5\simeq\Z_{10}$ as shown in App.~\ref{subsubsec:fAb}. In the edge theory, the fractional modes contribute the same $\Z_5$ zero mode sector as the $\nu=2/5$ states, while the additional integer quantum Hall mode contributes a $\Z_2$ Majorana sector. Together, the two sectors give the $\Z_{10}$ zero mode structure and a $20\pi$-periodic fractional Josephson effect.
\nocite{kapustin-2011,shen-2019,goddard-1985}
\bibliography{MR}
\end{document}